\documentclass[cite,a4paper,12pt]{article}
\usepackage{amsmath}
\usepackage{myart}
\usepackage{graphicx}
\usepackage{amsfonts}
\usepackage{amssymb}

\begin{document}
\hfill\hbox{RIKEN-TH-46}

\hfill\hbox{August 2005}

\begin{quotation}
The work is dedicated to A.Polyakov. I believe that

Sasha will recognize many of his ``low-dimensional''

ideas behind the present development.
\end{quotation}

\bigskip\ 

\begin{center}
{\Large \textbf{Perturbed Conformal Field Theory on Fluctuating Sphere}}

\vspace{1.5cm}

{\large Contribution to the Balkan Workshop BW2003, ``Mathematical,
Theoretical and Phenomenological Challenges Beyond Standard Model'', 29
August--02 September 2003, Vrnjancka Banja, Serbia}

\vspace{1.5cm}{\Large \ }

{\large Al.Zamolodchikov}\footnote{On leave of absence from Institute of
Theretical and Experimental Physics, B.Cheremushkinskaya 25, 117259 Moscow, Russia.}

\vspace{0.2cm}

Laboratoire de Physique Th\'eorique et Astroparticules\footnote{Laboratoire
Associ\'e au CNRS UMR 5207}

Universit\'e Montpellier II

Pl.E.Bataillon, 34095 Montpellier, France
\end{center}

\vspace{1.0cm}

\textbf{Abstract}

General properties of perturbed conformal field theory interacting with
quantized Liouville gravity are considered in the simplest case of spherical
topology. We discuss both short distance and large distance asymptotic of the
partition function. The crossover region is studied numerically for a simple
example of the perturbed Yang-Lee model, complemented in general with
arbitrary conformal ``spectator'' matter. The latter is not perturbed and
remains conformal along the flow, thus giving a control over the Liouville
central charge. The partition function is evaluated numerically from combined
analytic and perturbative information. In this paper we use the perturbative
information up to third order. At special points the four-point integral can
be evaluated and compared with our data. At the solvable point of minimal
Liouville gravity we are in remarkably good agreement with the matrix model
predictions. Possibilities to compare the result with random lattice
simulations is discussed.

\section{Introduction}

Liouville field theory (LFT) first appeared in the 2D quantum gravity context
in the famous paper by Polyakov \cite{Polyakov}. It has been demonstrated
there that the LFT action appears naturally as the induced action of a
conformal field theory living on a two-dimensional surface with non-trivilal
background metric. Quantum LFT was recognized as a conformal field theory
(CFT) and it is hardly an exaggeration to say that this peculiar example
always stayed behind the motivations in the subsequent explosive development
of the CFT, triggered by \cite{BPZ}. In particular, this induced LFT action
governs a specific dynamics of the 2D quantum gravity, which is now called the
critical (or Liouville) gravity. Characteristics of the critical gravity
(critical exponents, correlation functions etc.) depend on the central charge
of LFT, the last being simply related to the central charge of the inducing
conformal matter. As a conformal theory LFT\ is rather complicated and differs
in many important respects from the so called rational CFT's, developed and
studied soon after the pioneering paper \cite{BPZ}. Despite essential progress
in understanding of the space of states and hidden symmetries in LFT
(\cite{Thorn, Gervais} or \cite{D'Hoker}), physical interpretation of the
results in terms of fluctuating geometry of 2D surfaces remained unclear.

Meanwhile, a very different approach to 2D quantum gravity has been under
development. This approach can be called in general the discrete gravity. A
smooth Riemannian manifold is substituted by a ``random triangulation'', or
more precisely, by a graph of the same topology (and possibly certain
``matter'' degrees of freedom attached to its vertices or faces) while the
Riemann metric is simulated by the geodesic distances along the graph, the
total number of vertices (or plaquettes) being interpreted as the surface
area. The gravity problem is then reduced to a summation (in the spirit of the
path integral) over the ensemble of such graphs. Progress in the study of
these discrete models was drastically boosted after invention of a very
powerful method to sum up the statistical ensembles of discrete gravity by
reducing them to certain integrals over $N\times N$ matrices in the limit
$N\rightarrow\infty$ \cite{Kazakov, FDavid, KMK}. The latter integrals can be
sometimes evalutated explicitly, giving rise to exactly solvable models of
discrete gravity. Despite the fact that not every discrete model can be
treated by the matrix model approach, progress was so dramatic that now these
discrete models are often called the matrix models of 2D gravity. In the limit
$N\rightarrow\infty$ many types of critical behavior with different critical
exponents had been discovered. Moreover, it is often possible to calculate
explicitly the scaling functions which describe the ``flows'' between
different critical points. Another impressive advantage of the discrete
approach is that it allows to treat ``surfaces'' of arbitrary topology. This
led even to the study of the grand statistical ensemble of surfaces with
different topology, the object of extreme importance in the string theory.

Approximately at the same time the Polyakov's gravity has been reconsidered in
slightly different framework \cite{KPZ} (called sometimes the ``Polyakov's''
or ``light-cone'' gauge), which allowed to derive the critical exponents of
this induced gravity in an explicit form (the KPZ scaling). It was soon
recognized \cite{David, Distler} that the same exponents are exactly
reproduced in the original formulation of LFT provided certain finite
renormalization of the coupling constant are conjectured and taken into account.

The nexus of the Liouville and discrete approaches to quantum gravity in fact
flashed already in ref.\cite{KPZ} in the observation that some critical KPZ
exponents are precisely those appearing in the critical behavior of random
lattice systems.

Since the very advance of the discrete 2D gravity \cite{Kazakov, FDavid} in
1985, it always challenged the field theoretical community. The ambition is
not only to match with its achievements, i.e. learn to evaluate the
correlation numbers and the scaling functions (and also to work at higher
topologies). To reveal the field theory potential in 2D gravity to its full
extent means to go beyond the limits accessible by the matrix models. In this
paper I recall the most important steps and problems along this line. A
scaling function for the spherical partition function of a perturbed CFT
coupled to Liouville gravity is considered as a perturbative sum over the
correlation numbers. Although it is still a problem to evaluate directly the
whole set of these numbers, general arguments allow to establish important
analytic properties of the scaling function. In order, these analytic
properties restrict the scaling function to a large extent. In particular,
sometimes even a few first orders of the perturbative series allow to
reproduce the whole function, approximately but to a remarkably good accuracy.

To my opinion the main lesson to draw from this simple exercise is that the
Liouville gravity as formulated on the basis of early Polyakov's ideas is
indeed a consistent and pithy structure whose physical content is rich and
definitely extends beyond the relatively dull world offered by the matrix
models in their present formulation \cite{matrix} (recent refreshments can be
found in refs.\cite{All}).

\section{Perturbed Liouville gravity}

Liouville gravity (LG) is the 2D gravity induced by a ``critical matter''. Its
matter content is described by a certain ``matter'' CFT $\mathcal{M}%
_{\text{M}}$ with the central charge $c_{\text{M}}$. In the field theory
approach LG is described as a joint CFT $\mathcal{M}_{\text{M}}+$ Liouville
$+$ ghosts \cite{Polyakov}
\begin{equation}
\mathcal{L}_{\text{LG}}=\mathcal{L}_{\text{M}}+\mathcal{L}_{\text{L}%
}+\mathcal{L}_{\text{gh}}\label{LLG}%
\end{equation}
Conformal Liouville field theory is defined through the local Lagrangian
\begin{equation}
\mathcal{L}_{\text{L}}=\frac1{4\pi}(\partial_{a}\phi)^{2}+\mu e^{2b\phi
}\label{LLFT}%
\end{equation}
Its central charge $c_{\text{L}}$ is related to the parameter $b$ as
\begin{equation}
c_{\text{L}}=1+6Q^{2}\label{cL}%
\end{equation}
The combination
\begin{equation}
Q=b^{-1}+b\label{Q}%
\end{equation}
is called the background charge (basically for historical reasons). The scale
parameter $\mu$ is called the cosmological constant because in LFT
$\exp(2b\phi)d^{2}x$ is interpreted as the quantized area element of the 2D
space. The ghost CFT is the usual $BC$ system
\begin{equation}
\mathcal{L}_{\text{gh}}=\frac1\pi(C\bar\partial B+\bar C\partial\bar
B)\label{Lgh}%
\end{equation}
of spin $(2,-1)$ and central charge $c_{\text{gh}}=-26$ \cite{Polyakov}. The
gravity interpretation requires the total central charge of the joint CFT
(\ref{LLG}) to be $0$. In other words, parameter $b$ in (\ref{LLFT}) is
determined from the ``central charge balance'' equation
\begin{equation}
c_{\text{L}}+c_{\text{M}}=26\label{cbalance}%
\end{equation}

Presently we are going to consider the perturbed Liouville gravity where the
matter is not critical but described as a perturbed CFT
\begin{equation}
\mathcal{L}_{\text{matter}}=\mathcal{L}_{\text{M}}+\frac\lambda{2\pi}%
\Phi\label{Lmatter}%
\end{equation}
Here $\Phi$ is some relevant primary field from the spectrum of the matter CFT
of dimension $\Delta$ and $\lambda$ is the corresponding coupling constant. It
seems natural to take a simple solvable model of CFT as the unperturbed matter
field theory, e.g., one of the famous minimal models $M_{p/p^{\prime}}$
\cite{BPZ}. The perturbing field is then one of the relevant degenerate fields
of this minimal model and its correlation functions are either known
explicitly or constructible in principle. The matter central charge takes the
worldwide known value $c=c_{p,p^{\prime}}$ and therefore the Liouville
parameters are rigidly fixed by (\ref{cbalance}). E.g., if the matter CFT is
the free Majorana fermion CFT $\mathcal{M}_{1/2}$ the Liouville parameter is a
number $b=\sqrt{3/4}$ and no more a parameter. For example, a question like
``what is the physics of free fermion coupled to almost classic gravity?''
cannot be addressed. From theoretical point of view it seems convenient to
lift this restriction and get an independent control over the Liouville
parameters. To this order let us add to the matter content of the theory
another conformal matter of central charge $c_{\text{sp}}$. This matter is
unperturbed and always remain conformal. Therefore it formally does not
interact neither with the Liouville mode (except for through the conformal
anomaly) nor with the perturbed CFT (\ref{Lmatter}). Its main role is to
contribute to the central charge balance (\ref{cbalance}). For this reason we
call it the ``spectator'' matter. To be precise, there is also some connection
via conformal moduli. However in the simplest case of the spherical geometry
(which we mainly address presently) any details of this ``spectator'' CFT
except for its central charge are of no importance. It can equally well be
considered as a formal shapeless ingredient seen only through $c_{\text{sp}}$.
In the last section we'll discuss possible realizations of the spectator matter.

With this additional liberty the total matter central charge is
\begin{equation}
c_{\text{M}}=c+c_{\text{sp}}\label{cM}%
\end{equation}
and the perturbed Liouville gravity is expressed symbolically through the
Lagrangian
\begin{equation}
\mathcal{L}_{\lambda}=\mathcal{L}_{\text{CFT}}+\mathcal{L}_{\text{L}}%
+\frac\lambda{2\pi}\Phi e^{2g\phi}+\mathcal{L}_{\text{sp}}+\mathcal{L}%
_{\text{gh}}\label{pLG}%
\end{equation}
As explained above, in the spherical case the last two terms in (\ref{pLG})
are coupled only through the conformal anomaly. Therefore we will omit these
background degrees of freedom, having in mind that they enter the central
charge balance, which is solved for $b$ as
\begin{equation}
b=\sqrt{\frac{25-c_{\text{M}}}{24}}-\sqrt{\frac{1-c_{\text{M}}}{24}}\label{bc}%
\end{equation}
For the branch chosen the semiclassical regime $b\rightarrow0$ is achieved
when $c_{\text{sp}}\rightarrow-\infty$. In this limit the background space is
in fact a classical symmetric sphere. Otherwise this parameter $b$
characterizes the ``rigidity'' of the space with respect to quantum fluctuations.

Interaction with the Liouville gravity ``dresses'' the perturbing primary
field $\Phi$ with an appropriate Liouville exponential field $\exp(2g\phi) $.
Parameter $g$ is determined from the following kinematic requirement
\begin{equation}
g(Q-g)+\Delta=1\label{Dbalance}%
\end{equation}
which means simply that the composite field
\begin{equation}
U=\Phi e^{2g\phi}\label{Ug}%
\end{equation}
is a $(1,1)$ form and can be integrated over the space. Ordinarily the least
of the two roots is chosen
\begin{equation}
g=\sqrt{\frac{25-c_{\text{M}}}{24}}-\sqrt{\frac{1-c_{\text{M}}}{24}+\Delta
}\label{g}%
\end{equation}
Notice that $g$ is positive for relevant perturbations $\Delta<1$ (and
negative for the irrelevant ones).

\section{Spherical partition function}

The spherical partition function $Z(\mu,0)$ of the Liouville gravity is known
to scale as
\begin{equation}
Z(\mu,0)=\mu^{Q/b}Z(1,0)\label{Zmu}%
\end{equation}
with some dimensionless constant $Z(1,0)$. It is easy to see that the
perturbing coupling constant $\lambda$ scales as $\mu^{1/\rho}$ with positive
\begin{equation}
\rho=bg^{-1}\label{s}%
\end{equation}
The perturbed partition function $Z(\mu,\lambda)$ introduces the scaling
function $\mathcal{Z}$ of the dimensionless combination $\lambda\mu^{-1/\rho
}$
\begin{equation}
Z(\mu,\lambda)=\mu^{Q/b}\mathcal{Z}\left(  \lambda\mu^{-1/\rho}\right)
\label{Zscale}%
\end{equation}

This scaling function is the main object of our present study. Sometimes it
will be more convenient to consider the ``fixed area'' partition function
$Z_{A}(\lambda)$ defined as the inverse Laplace transform of $Z(\mu,\lambda)$%
\begin{equation}
Z_{A}(\lambda)=A\int_{\uparrow}\frac{d\mu}{2\pi i}Z(\mu,\lambda)e^{\mu
A}\label{ZAlambda}%
\end{equation}
where the contour $\uparrow$ goes along the imaginary axis to the right from
all singularities of the integrand. In particular, in the critical gravity
\begin{equation}
Z_{A}(0)=Z(1,0)\int_{\uparrow}\frac{d\mu}{2\pi i}\mu^{Q/b}e^{\mu A}%
=\frac{Z(1,0)}{\Gamma(-Qb^{-1})}A^{-Q/b}\label{ZA0}%
\end{equation}
Inversely
\begin{equation}
Z(\mu,\lambda)=\int_{(0)}^{\infty}Z_{A}(\lambda)e^{-\mu A}\frac{dA}%
A\label{Laplace}%
\end{equation}
where $(0)$ at the lower limit indicates that the integral is often divergent
at small $A$ and (\ref{Laplace}) means a properly regularized version. The
name ``fixed area'' is motivated by the interpretation of the integral
\[
A=\int e^{2b\phi}d^{2}x
\]
as the total area of the gravitating sphere. For the same reasons as for
$Z(\mu,\lambda)$ the fixed area partition function has similar scaling form
\begin{equation}
Z_{A}(\lambda)=Z_{A}(0)z(h)\label{zh}%
\end{equation}
where it happens more convenient to normalize the scale invariant parameter
$h$ as
\begin{equation}
h=\lambda\left(  \frac A\pi\right)  ^{1/\rho}\label{h}%
\end{equation}

\section{UV expansion}

Partition function (\ref{Zscale}) is a regular series in the coupling constant
$\lambda$
\begin{equation}
\mathcal{Z}(\mu,\lambda)=\sum_{n=0}^{\infty}a_{n}\left(  -\lambda\right)
^{n}\label{Zan}%
\end{equation}
Each term in this expansion is determined by the corresponding order in the
interaction (\ref{Ug}). The coefficients
\begin{equation}
a_{n}=\frac{Z^{-1}(\mu,0)}{(2\pi)^{n}n!}\int_{M_{n}}\left\langle
U^{n}\right\rangle \label{an}%
\end{equation}
are the integrated (and normalized) correlation functions of $n$ insertions of
the composite (\ref{Ug}). According to (\ref{Zscale}) the coefficients $a_{n}$
scale as $a_{n}\sim\mu^{-n/\rho}$. In the LFT approach the source correlation
functions (i.e., the coordinate dependent functions before the integration)
factorize into a product
\begin{equation}
\left\langle U(x_{1})\ldots U(x_{n})\right\rangle =\left\langle \Phi
(x_{1})\ldots\Phi(x_{n})\right\rangle _{\text{M}}\left\langle e^{2g\phi
(x_{1})}\ldots e^{2g\phi(x_{n})}\right\rangle _{\text{L}}\label{LFact}%
\end{equation}
as it can be literally read off from the action (\ref{pLG}). The integral in
(\ref{an}) is not over the positions of all $n$ insertions but rather over the
$n-3$ dimensional moduli space $M_{n}$ of the sphere with $n$ punctures.
Practically at $n\geq3$ this amounts to place any $3$ insertions of $U$ at
arbitrary fixed points $x_{1}$, $x_{2}$ and $x_{3}$ and integrate over the
positions of the remaining $n-3$ ones
\begin{equation}
\int_{M_{n}}\left\langle U^{n}\right\rangle =\int\left\langle W(x_{1}%
)W(x_{2})W(x_{3})U(x_{4})\ldots U(x_{n)}\right\rangle d^{2}x_{4}\ldots
d^{2}x_{n}\label{Un}%
\end{equation}
Following the standard string theory practice we have also replaced the
``nailed'' insertions $U(x_{1})$ by $W(x_{i})=C\bar CU(x_{i})$ to get rid of
the redundant dependence on their positions and also to give the integrand the
total ghost number $3$ (as it is required by the ghost number conservation).
Simplest example is the three-point function, where the moduli space is empty
and the three-point function is simply the product
\begin{equation}
\left\langle U^{3}\right\rangle =\left\langle C\bar C(x_{1})C\bar
C(x_{2})C\bar C(x_{3})\right\rangle _{\text{gh}}\left\langle \Phi(x_{1}%
)\Phi(x_{2})\Phi(x_{3})\right\rangle _{\text{M}}\left\langle e^{2g\phi(x_{1}%
)}e^{2g\phi(x_{2})}e^{2g\phi(x_{3})}\right\rangle _{\text{L}}\label{U3}%
\end{equation}
Two- and zero-point functions are simply figured out from the three-point one
(see below).

In the exactly solvable CFT models all multipoint correlation functions are in
principle constructible. They are determined uniquely once the normalization
of the primary fields is fixed. In what follows we use the standard CFT
normalization of the matter fields through the two-point function
\begin{equation}
\left\langle \Phi(x_{1})\Phi(x_{2})\right\rangle _{\text{M}}=\left|
x_{1}-x_{2}\right|  ^{-4\Delta}\label{Phi2}%
\end{equation}
The field normalization also gives precise meaning to the coupling constant
$\lambda$ in (\ref{Lmatter}) and (\ref{pLG}).

In practice, even in solvable CFT's, it is usually difficult to go explicitly
beyond the four-point functions (except for such special cases as the
free-field CFT's). At the same time the one-, two- and three-point functions
are rather universal. Conventional CFT wisdom requires the one point function
to vanish
\begin{equation}
\left\langle \Phi(x)\right\rangle _{\text{M}}=0\label{Phi0}%
\end{equation}
while the three-point function is fixed up to an overall constant $C_{\Phi
\Phi\Phi}$%
\begin{equation}
\left\langle \Phi(x_{1})\Phi(x_{2})\Phi(x_{3})\right\rangle _{\text{M}}%
=\frac{C_{\Phi\Phi\Phi}}{\left|  x_{12}x_{23}x_{32}\right|  ^{2\Delta}%
}\label{Phi3}%
\end{equation}
which is called the three-$\Phi$ structure constant and is usually known
explicitly in solvable CFT's. In (\ref{Phi3}) and often below we use the
abbreviated notation $x_{kl}=x_{k}-x_{l}$. Higher correlation functions have
less universal form. The four-point case is more handy and allows explicit
channel decompositions, e.g.,
\begin{align}
\  &  \left\langle \Phi(x_{1})\Phi(x_{2})\Phi(x_{3})\Phi(x_{4})\right\rangle
_{\text{M}}=\frac{G_{\text{M}}(x,\bar x)}{\ \left|  x_{14}x_{23}\right|
^{4\Delta}}\label{Phi4}\\
G_{\text{M}}(x,\bar x) &  =\sum_{i}C_{\Phi\Phi}^{i}C_{\Phi\Phi i}%
\mathcal{F}\left(
\begin{array}
[c]{cc}%
\Delta & \Delta\\
\Delta & \Delta
\end{array}
\left|  \Delta_{i}\right|  x\right)  \mathcal{F}\left(
\begin{array}
[c]{cc}%
\Delta & \Delta\\
\Delta & \Delta
\end{array}
\left|  \Delta_{i}\right|  \bar x\right) \nonumber
\end{align}
in terms of the structure constants and the four-point conformal blocks
related to the conformal algebra with central charge $c$ \cite{BPZ}
\begin{equation}
\mathcal{F}\left(
\begin{array}
[c]{cc}%
\Delta & \Delta\\
\Delta & \Delta
\end{array}
\left|  \Delta_{i}\right|  x\right)  =%
\raisebox{-0.4583in}{\includegraphics[
height=1.0205in,
width=2.1352in
]%
{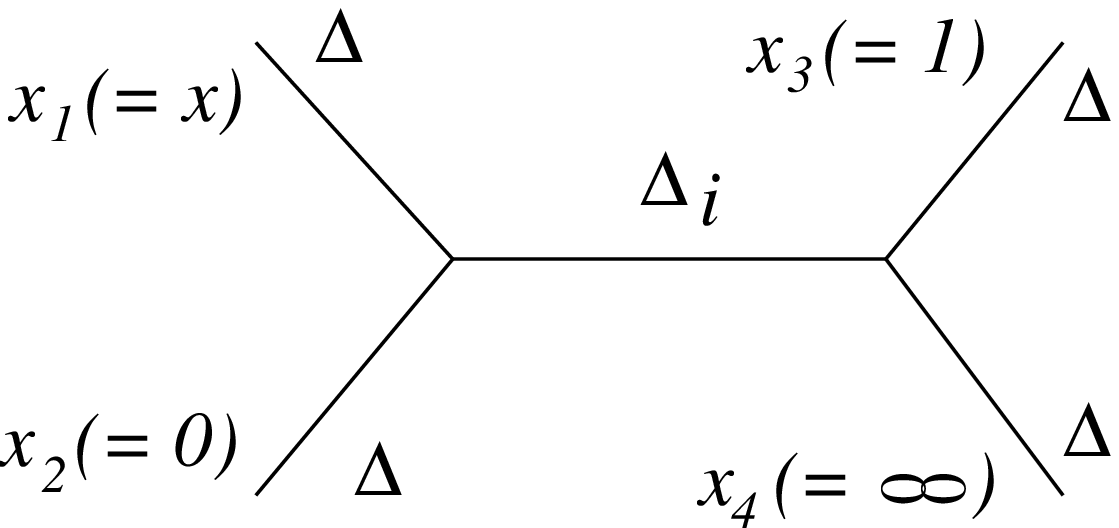}%
}%
\label{Mblock}%
\end{equation}
These blocks are holomorphic in the projective invariant
\begin{equation}
x=\frac{(x_{1}-x_{2})(x_{3}-x_{4})}{(x_{1}-x_{4})(x_{3}-x_{2})}\label{x}%
\end{equation}
The sum in (\ref{Phi4}) is over all primary fields $\left\{  \Phi_{i}%
,\Delta_{i}\right\}  $ entering the operator product expansion of $\Phi
(x)\Phi(0)$.

The situation is nearly the same in LFT. The three-point functions are known
explicitly \cite{DO} for arbitrary exponential operators $\exp(2a\phi)$ (the
basic primary fields of LFT)
\begin{equation}
\left\langle e^{2a_{1}\phi(x_{1})}e^{2a_{2}\phi(x_{2})}e^{2a_{3}\phi(x_{3}%
)}\right\rangle _{\text{L}}=\frac{C_{\text{L}}(a_{1},a_{2},a_{3})}{\left|
x_{12}\right|  ^{2\Delta_{1}+2\Delta_{2}-2\Delta_{3}}\left|  x_{23}\right|
^{2\Delta_{3}+2\Delta_{4}-2\Delta_{1}}\left|  x_{31}\right|  ^{2\Delta
_{3}+2\Delta_{1}-2\Delta_{2}}}\label{V3}%
\end{equation}
where
\begin{equation}
\Delta_{i}=a_{i}(Q-a_{i})\label{DL}%
\end{equation}
are the dimensions of the LFT exponential fields and the constant (letter $a$
below stands for $a_{1}+a_{2}+a_{3}$)
\begin{equation}
\ C_{\text{L}}(a_{1},a_{2},a_{3})=\left(  \pi\mu\gamma(b^{2})b^{2-2b^{2}%
}\right)  ^{(Q-a)/b}\frac{\Upsilon_{b}(b)}{\Upsilon_{b}(a-Q)}\prod_{i=1}%
^{3}\frac{\Upsilon_{b}(2a_{i})}{\Upsilon_{b}(a-2a_{i})}\label{CL}%
\end{equation}
is expressed in terms of certain special function $\Upsilon_{b}(x)$, which is
an entire function related to the Barnes double gamma function. It's defining
properties are the shift relations
\begin{align}
\Upsilon_{b}(x+b) &  =\gamma(bx)b^{1-2bx}\Upsilon_{b}(x)\label{Yb}\\
\Upsilon_{b}(x+b^{-1}) &  =\gamma(b^{-1}x)b^{-1+2b^{-1}x}\Upsilon
_{b}(x)\nonumber
\end{align}
It is also symmetric w.r.t. the reflection $x\rightarrow Q-x$ and self-dual
(i.e., invariant w.r.t. $b\rightarrow b^{-1}$) as a function of the parameter
$b$. Integral representation, which fixes also the overall normalization can
be found e.g. in \cite{AAl}. In (\ref{CL}) and throughout the paper
$\gamma(x)$ is the usual little $\gamma$-function. The LFT four-point
function
\begin{align}
\  &  \left\langle e^{2a_{1}\phi(x_{1})}e^{2a_{2}\phi(x_{2})}e^{2a_{3}%
\phi(x_{3})}e^{2a_{4}\phi(x_{4})}\right\rangle _{\text{LFT}}=\label{V4}\\
&  \ \ \frac{\left|  x_{24}\right|  ^{2(\Delta_{1}+\Delta_{3}-\Delta
_{2}-\Delta_{4})}\left|  x_{34}\right|  ^{2(\Delta_{1}+\Delta_{2}-\Delta
_{3}-\Delta_{4})}}{\left|  x_{14}\right|  ^{4\Delta_{1}}\left|  x_{23}\right|
^{2(\Delta_{1}+\Delta_{2}+\Delta_{3}-\Delta_{4})}}\ G_{\text{L}}\left(
\left.
\begin{array}
[c]{cc}%
a_{1} & a_{3}\\
a_{2} & a_{4}%
\end{array}
\right|  x,\bar x\right) \nonumber
\end{align}
also allows holomorphic antiholomorphic decomposition \cite{AAl}, similar to
(\ref{Phi4})
\begin{align}
\  &  G_{\text{L}}\left(  \left.
\begin{array}
[c]{cc}%
a_{1} & a_{3}\\
a_{2} & a_{4}%
\end{array}
\right|  x,\bar x\right)  =\label{V4block}\\
&  \ \ \ \ \ \int\frac{dP}{4\pi}C_{\text{L}}\left(  a_{1},a_{2},\frac
Q2+iP\right)  C_{\text{L}}\left(  \frac Q2-iP,a_{3},a_{4}\right)
\mathcal{F}_{P}\left(  \left.
\begin{array}
[c]{cc}%
\Delta_{1} & \Delta_{3}\\
\Delta_{2} & \Delta_{4}%
\end{array}
\right|  x\right)  \mathcal{F}_{P}\left(  \left.
\begin{array}
[c]{cc}%
\Delta_{1} & \Delta_{3}\\
\Delta_{2} & \Delta_{4}%
\end{array}
\right|  \bar x\right) \nonumber
\end{align}
In (\ref{V4}) and (\ref{V4block}) $\Delta_{i}$ with $i=1,\ldots,4$ are the LFT
dimensions (\ref{DL}).The integral now is over a continuous spectrum of
irreducible representations of the Virasoro algebra with central charge
$c_{\text{L}}$ and dimension $Q^{2}/4+P^{2}$. The corresponding four-point
blocks
\begin{equation}
\mathcal{F}_{P}\left(  \left.
\begin{array}
[c]{cc}%
\Delta_{1} & \Delta_{3}\\
\Delta_{2} & \Delta_{4}%
\end{array}
\right|  x\right)  =%
\raisebox{-0.4722in}{\includegraphics[
height=1.0395in,
width=2.2719in
]%
{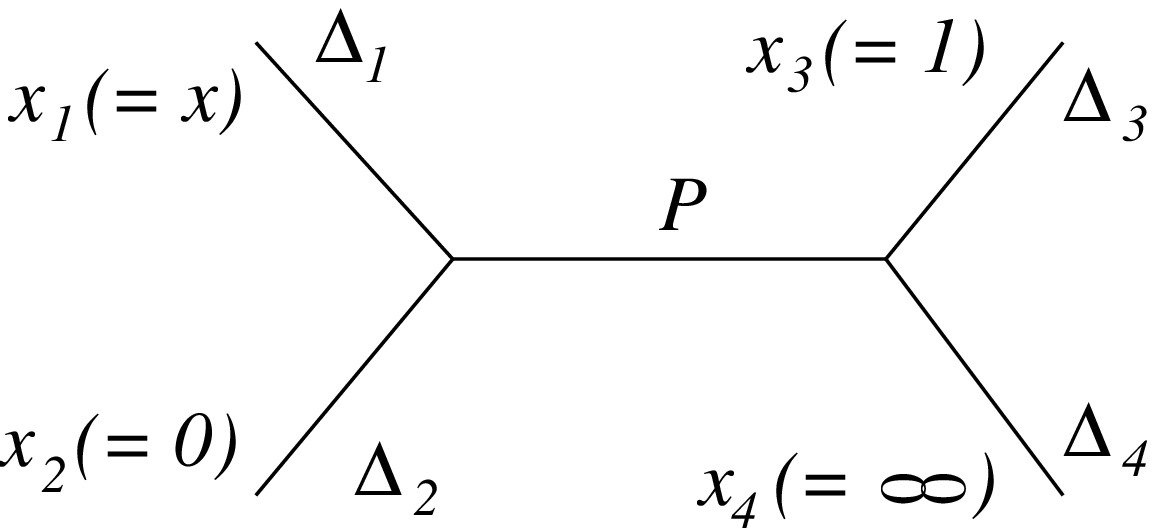}%
}%
\label{Lblock}%
\end{equation}
are related to the conformal algebra with the Liouville central charge
$c_{\text{L}}$.

Application of eq.(\ref{Un}) is not straightforward at $n<3$. The moduli space
has formally negative dimension and this is related to the non-trivial space
of conformal Killing vector fields on the sphere with $2$ and $0$ punctures.
The simplest way to determine the correlation functions with $n<3$ is to
differentiate them sufficiently many times with respect to $\mu$ (each
differentiation introduces an insertion of the area-element field $\exp
(2b\phi)$) until the order of the correlation function attains $n\geq3 $.This
maneuver is quite a common way to treat the zero-mode problem of the sphere.
In the case of the two-point function it's enough to differentiate once
\begin{equation}
-\frac\partial{\partial\mu}\left\langle e^{2g\phi}e^{2g\phi}\right\rangle
_{\text{L}}=C_{\text{L}}(g,g,b)=\left(  \pi\mu\gamma(b^{2})\right)
^{(Q-2g)/b}\frac{\gamma(2gb-b^{2})\gamma(2gb^{-1}-b^{-2})}{\pi\mu
b}\label{dV2}%
\end{equation}
After an integration in $\mu$%
\begin{equation}
\left\langle e^{2g\phi}e^{2g\phi}\right\rangle _{\text{L}}=-\left(  \pi
\mu\gamma(b^{2})\right)  ^{(Q-2g)/b}\frac{\gamma(2gb-b^{2})\gamma
(2gb^{-1}-b^{-2})}{\pi(Q-2g)}\label{V2}%
\end{equation}
This expression \emph{does not }coincide with the Liouville reflection
amplitude \ref{AAl}
\begin{equation}
R(g)=-\left(  \pi\mu\gamma(b^{2})\right)  ^{(Q-2g)/b}\frac{\gamma
(2gb-b^{2})\gamma\left(  2gb^{-1}-b^{-2}\right)  }{\left(  2g-Q\right)  ^{2}%
}\label{Rg}%
\end{equation}
(which is ordinarily associated with the two-point function in LFT) differing
in the factor of $\pi^{-1}(Q-2g)$. Similar wisdom gives for the Liouville
partition function
\begin{equation}
-\frac{\partial^{3}}{\partial\mu^{3}}Z_{\text{L}}=C_{\text{L}}(b,b,b)=-\left(
\pi\mu\gamma(b^{2})\right)  ^{Q/b}\frac{\left(  b^{-2}-1\right)  ^{2}}{\pi
^{3}\mu^{3}b\gamma(b^{2})\gamma\left(  b^{-2}\right)  }\label{d3ZL}%
\end{equation}
and hence
\begin{equation}
Z_{\text{L}}=\left(  \pi\mu\gamma(b^{2})\right)  ^{Q/b}\frac{(1-b^{2})}%
{\pi^{3}Q\gamma(b^{2})\gamma(b^{-2})}\label{ZL}%
\end{equation}
Notice that this expression in \emph{not }self-dual, unlike the three-point
function itself. This fact looks somewhat surprising in the Liouville context
and likely needs better understanding. Below we will meet many opportunities
to support the consistency of (\ref{V2}) and (\ref{ZL}) in the 2D gravity
context. Therefore we have to take this breakdown of the Liouville
self-duality in the 2D gravity application as the matter of fact.

After all these preliminaries, we are in the position to construct explicitly
few first terms in the perturbative expansion (\ref{Zan})%

\begin{align}
a_{0}  &  =1\nonumber\\
a_{1}  &  =0\label{a1234}\\
a_{2}  &  =\left(  \pi\mu\gamma(b^{2})\right)  ^{-2g/b}\frac{Q\gamma
(2gb-b^{2})\gamma(2gb^{-1}-b^{-2})\gamma(b^{2})\gamma(b^{-2})}{8(2g-Q)(1-b^{2}%
)}\nonumber\\
a_{3}  &  =\frac{C_{\Phi\Phi\Phi}C_{\text{L}}(g,g,g)}{48\pi^{3}Z_{\text{L}}%
}\nonumber\\
a_{4}  &  =\frac1{24(2\pi)^{4}Z_{\text{L}}}\int G_{\text{M}}(x,\bar
x)G_{\text{L}}\left(  \left.
\begin{array}
[c]{cc}%
g & g\\
g & g
\end{array}
\right|  x,\bar x\right)  d^{2}x\nonumber\\
&  \ldots\nonumber
\end{align}

In what follows we will be more interested in the fixed area partition
function (\ref{zh}). The scaling function $z(h)$ has the following
perturbative expansion
\begin{equation}
z(h)=\sum_{n=0}^{\infty}z_{n}(-h)^{n}\label{zzn}%
\end{equation}
where the dimensionless coefficients $z_{n}$ are related to $a_{n}$ as
\begin{equation}
z_{n}=\frac{(\pi\mu)^{n/\rho}a_{n}\Gamma(-1-b^{-2})}{\Gamma(n\rho^{-1}%
-b^{-2}-1)}\label{zn}%
\end{equation}
Explicitly up to the third order
\begin{align}
z_{0} &  =1\nonumber\\
z_{1} &  =0\label{z123}\\
z_{2} &  =\left(  \gamma(b^{2})\right)  ^{-2g/b}\frac{\gamma(2gb-b^{2}%
)\gamma(b^{2})\Gamma(b^{-2})}{8(b^{-2}-1)\Gamma(1+b^{-2}-2gb^{-1})}\nonumber\\
z_{3} &  =\frac{C_{\Phi\Phi\Phi}}{6(2\pi)^{3}}\frac{C_{\text{L}}^{(A=\pi
)}(g,g,g)}{Z_{\text{L}}^{(A=\pi)}}\nonumber
\end{align}
In the last expression
\begin{equation}
C_{\text{L}}^{(A)}(g,g,g)=\frac{\left(  \pi\gamma(b^{2})b^{2-2b^{2}}%
A^{-1}\right)  ^{(Q-3g)/b}\Upsilon(b)\Upsilon^{3}(2g)}{\Upsilon(3g-Q)\Upsilon
^{3}(g)\Gamma(3gb^{-1}-1-b^{-2})}\label{C3A}%
\end{equation}
is the fixed area $A$ Liouville three-point function and the fixed area
partition sum $Z_{\text{LFT}}^{(A)}$ reads
\begin{equation}
Z_{\text{L}}^{(A)}=\left(  \frac{\pi\gamma(b^{2})}A\right)  ^{Q/b}\frac
{\Gamma(2-b^{2})}{\pi^{3}b^{3}\Gamma(b^{2})\Gamma(b^{-2})}\label{ZA}%
\end{equation}

For relevant perturbations (\ref{h}) is small for small area $A$. This means
that the perturbative expansion (\ref{zzn}) in fact determines the behavior of
the partition function at small surface area $A$. Notice also that the
coefficients $z_{n}$ in (\ref{zn}) are suppressed at large $n$ by additional
$\Gamma$-function in the denominator as compared with $a_{n}$. This means that
if the perturbative series (\ref{Zan}) is convergent, the fixed area $z(h)$ is
an entire function of $h$.

\section{Criticality}

From physical considerations we expect that at $A\rightarrow\infty$ the fixed
area partition function behaves as
\begin{equation}
Z_{A}(\lambda)=Z_{\infty}A^{-Q^{\prime}/b^{\prime}}e^{\mu_{\text{c}}%
A}\label{ZAinf}%
\end{equation}
where $\mathcal{E}_{\text{vac}}(b)=-\mu_{\text{c}}$ is the specific free
energy generated by the massive (finite correlation length) modes of the
perturbed CFT $\mathcal{M}_{c}$. For dimensional reasons
\begin{equation}
\mu_{\text{c}}=f_{0}(b)\lambda^{\rho}\label{f0}%
\end{equation}
with some dimensionless $f_{0}(b)$. This perturbed CFT interacts also with the
quantized gravity and therefore $\mathcal{E}_{\text{vac}}(b)$ reduces to the
vacuum energy $\mathcal{E}_{\text{vac}}(0)$ of the model (\ref{Lmatter}) in
``rigid'' flat background only in the classical limit $b^{2}=0$. Otherwise it
depends on $b^{2}$ and appears to be an important characteristic of the matter
interacting with quantum gravity. Clearly the grand partition function
$Z(\mu,\lambda)$ develops a singularity at $\mu=\mu_{\text{c}}$. Therefore
$\mathcal{E}_{\text{vac}}(b)$ is simply another interpretation of the critical
point in the cosmological constant (usually referred to as the value of $\mu$
where the surface ``blows up'', one of the images convenient to make an
illusion of understanding). The related singularity of $Z(\mu,\lambda)$ in the
coupling constant $\lambda$ at
\begin{equation}
\lambda_{\text{c}}=\left(  \frac\mu{f_{0}(b)}\right)  ^{1/\rho}\label{lcrit}%
\end{equation}
often happens to be the one closest to the origin. In this case it determines
the finite convergence region of the perturbative series (\ref{Zan}). As it
was already mentioned above, this in order implies that the function $z(h)$ is entire.

Let us also assume that the perturbed matter is ``massive'', i.e., the
correlations of this coupled matter degrees of freedom do not extend far
beyond the characteristic mass scale $m_{\text{c}}\sim$ $\lambda^{\rho}$. Then
at $A\gg m_{\text{c}}^{-2}$ the ``massive'' part of the matter ``dies out'',
i.e., it does not contribute any more to the matter central charge. The
gravity dynamics is determined then by the central charge of the ``spectator
matter'', which doesn't interact and therefore always remains conformal. Then
the infrared parameters $b^{\prime}$ and
\begin{equation}
Q^{\prime}=1/b^{\prime}+b^{\prime}\label{Qprim}%
\end{equation}
are determined from the equation
\begin{equation}
1+6(Q^{\prime})^{2}+c_{\text{sp}}=26\label{IRbalance}%
\end{equation}
These are the parameters entering the power-like part of the asymptotic
(\ref{ZAinf}). More explicitly
\begin{align}
Q^{\prime} &  =\sqrt{Q^{2}+\frac c6}\label{Qprime}\\
b^{\prime} &  =\sqrt{\frac{Q^{2}}4+\frac c{24}}-\sqrt{\frac{Q^{2}}4+\frac
c{24}-1}\nonumber
\end{align}

In terms of $h$ the asymptotic of the function reads
\begin{equation}
\log z(h)\sim\pi f_{0}(b)h^{\rho}+\rho\left(  \frac Qb-\frac{Q^{\prime}%
}{b^{\prime}}\right)  \log h+\log z_{\infty}+\ldots\label{logzh}%
\end{equation}
where we have also added an undeterminate constant term. Dots stand for the
subsequent decreasing terms in the large $h$ expansion. In particular $\rho$
is the order of the entire function $z(h)$.

Under certain additional assumptions about the analytic properties of $z(h)$
the complementary information provided by the asymptotic (\ref{logzh}) and the
first few terms in the series expansion (\ref{zzn}) turns out quite
restrictive and permits to achieve rather detailed description of this
function. The efficiency of the combined analytic-numeric treatment
implemented below depends decisively on these additional assumptions, which do
not always hold true, being dependent on the choice of the matter field
theory. From now on we restrict ourselves to a very particular example of the
perturbed CFT, the scaling Yang-Lee model, where these additional properties
are most likely valid. Therefore the central charge $c$ in (\ref{cM}) and the
perturbing dimension $\Delta$ in (\ref{Dbalance}) are now fixed $c=-22/5$,
$\Delta=-1/5$. The ``spectator'' matter however remains undefined, the
Liouville coupling $b$ being a variable parameter of the model.

\section{Scaling Yang-Lee model}

The scaling Yang-Lee model \cite{SLYM} is often used as a testing tool for
various approximate methods in 2D field theory. It is probably the simplest
(after the free field theories) example of perturbed CFT. The model arises as
a perturbation of the non-unitary minimal model $\mathcal{M}_{2/5}$. This CFT
is rational and contains only two primary fields, the identity $I$ of
dimension $0$ and the basic field $\varphi=\Phi_{1,3}$ of dimension
$\Delta=-1/5$. The basic operator product expansion reads
\begin{equation}
\varphi(x)\varphi(0)=(x\bar x)^{2/5}\left[  I\right]  +(x\bar x)^{1/5}%
C_{\varphi\varphi\varphi}\left[  \varphi(0)\right] \label{ppp}%
\end{equation}
where the three-$\varphi$ structure constant is purely imaginary
$C_{\varphi\varphi\varphi}=i\kappa$ with
\begin{equation}
\kappa=\frac{\gamma^{3/2}(1/5)}{5\gamma^{1/2}(3/5)}=1.91131269990474\ldots
\label{kappa}%
\end{equation}
Below we will need also the matter four point function (\ref{Phi4})
\begin{equation}
\left\langle \varphi(x)\varphi(0)\varphi(1)\varphi(\infty)\right\rangle
_{\text{YL}}=G_{\text{YL}}(x,\bar x)=\mathcal{F}_{I}(x)\mathcal{F}_{I}(\bar
x)-\kappa^{2}\mathcal{F}_{\varphi}(x)\mathcal{F}_{\varphi}(\bar x)\label{GYL}%
\end{equation}
where the blocks are explicitly in terms of hypergeometric functions
\begin{align}
\mathcal{F}_{I}(x) &  =x^{2/5}(1-x)^{1/5}{}_{2}F_{1}%
(2/5,3/5,6/5,x)\label{YLblocks}\\
\mathcal{F}_{\varphi}(x) &  =x^{1/5}(1-x)^{1/5}{}_{2}F_{1}%
(1/5,2/5,4/5,x)\nonumber
\end{align}

The only relevant perturbation by the field $\varphi$ gives rise to what is
called the scaling Yang-Lee model. To make the theory real the corresponding
coupling constant has to be taken pure imaginary. With this choice in
(\ref{pLG})
\begin{equation}
\mathcal{L}_{\text{matter}}=\mathcal{L}_{\text{YL}}+\frac{i\lambda}{2\pi
}\varphi(x)e^{2g\phi}+\mathcal{L}_{\text{sp}}\label{ASLYM}%
\end{equation}
where $g$ is determined by (\ref{Dbalance}). In a flat classical space-time
this model is integrable. Its particle content and factorized scattering
theory are known exactly \cite{CM}. The spectrum contains only one massive
particle of mass $m$. Integrability provides us with a number of exact
results. E.g., the exact relation
\begin{equation}
m=k\lambda^{5/12}\label{mlambda}%
\end{equation}
between the mass $m$ and the coupling constant $\lambda$ is known
\cite{masscale}
\begin{equation}
k=\frac{4\sqrt{\pi}}{\Gamma(5/6)\Gamma(2/3)}\left[  \frac{\Gamma
(4/5)\Gamma(3/5)}{100\Gamma(1/5)\Gamma(2/5)}\right]  ^{5/24}%
=1.2288903248\ldots\label{mr}%
\end{equation}
The bulk vacuum energy is also obtained explicitly \cite{TBA, DeVega}%

\begin{equation}
\mathcal{E}_{\text{vac}}=-\frac{m^{2}}{4\sqrt{3}}\label{ESLYM}%
\end{equation}
The classical space-time background is achieved in the limit $b=0$. Hence the
dimensionless characteristic $f_{0}(b^{2})$ in (\ref{f0}) is known exactly at
$b^{2}=0$
\begin{equation}
f_{0}(0)=\frac{k^{2}}{4\sqrt{3}}=0.2179745\ldots\label{f00}%
\end{equation}

\section{Analytic-numeric treatment}

The analytic-numeric treatment here altogether follows the lines of
ref.\cite{sphere}, where precisely the same scaling Yang-Lee model has been
considered on the rigid classical sphere. This rigid spherical background in
our present case corresponds to the classical limit $b^{2}=0$. In this sense
the present work is a direct extension of \cite{sphere} for the quantized geometry.

In \cite{sphere} it was observed that the asymptotic (\ref{logzh}) holds in
the whole complex plane of $h$ except for the negative real axis. All the
zeros $h_{n}$, $n=0,1,\ldots$ of $z(h)$ are real, all but the first one
$h_{0}>0 $ being negative. In the present case of quantum geometry we make an
assumption that these properties still hold for the whole allowed domain of
the parameter $0\leq b^{2}\leq b_{\text{c}}^{2}$. The upper limit here
\begin{equation}
b_{\text{c}}^{2}=\frac{7-2\sqrt{6}}5=0.4202041028\ldots\label{bcrit}%
\end{equation}
is related to the fact that at this value the ``dressing'' parameter $g$ in
(\ref{Dbalance}) attains its critical value $g=Q/2$. For $b^{2}>b_{\text{c}%
}^{2}$ the solutions to (\ref{Dbalance}) are complex. The assumed location of
the zeros is supported by the consistency of the numeric analysis of the next
section. The presumption made, the asymptotic along the negative real axis
reads (for brevity we suppress the argument $b$ in the free energy
$f_{0}=f_{0}(b)$)
\begin{equation}
z(h)\sim\exp\left(  \pi f_{0}\cos\left(  \pi\rho\right)  (-h)^{\rho}%
+\ldots\right)  \sin\left(  \pi f_{0}\sin\left(  \pi\rho\right)  (-h)^{\rho
}+\pi\delta+\ldots\right) \label{asneg}%
\end{equation}
where
\begin{equation}
\delta=\rho\left(  \frac Qb-\frac{Q^{\prime}}{b^{\prime}}\right)
-\frac12\label{bzdel}%
\end{equation}
The asymptotic location of zeros $h_{n}$ at $n\rightarrow\infty$ is hence
estimated as
\begin{equation}
-h_{n}\sim\left(  \frac{n-\delta}{f_{0}\sin(\pi\rho)}\right)  ^{1/\rho
}\label{hass}%
\end{equation}
It is also convenient, as in \cite{sphere}, to introduce the $\zeta$-function
of these zeros
\begin{equation}
\zeta_{\text{YL}}(s)=\frac{e^{i\pi s}}{h_{0}^{s}}+\sum_{n=1}^{\infty}%
\frac1{(-h_{n})^{s}}\label{zetaYL}%
\end{equation}
The scaling function itself $z(h)$ is related to (\ref{zetaYL}) through
\begin{equation}
\log z(h)=\int_{\uparrow}\frac{\pi\zeta_{\text{YL}}(s)}{\sin(\pi s)}h^{s}%
\frac{ds}{2i\pi s}\label{melline}%
\end{equation}
where the contour goes along the imaginary axis to the left from the pole at
$s=1$ and to the right of the pole at $s=\rho$ (note that $\rho<1$ for the
whole family of gravitational Yang-Lee models, see table \ref{Table2} below).
Inversely
\begin{equation}
\zeta_{\text{YL}}(s)=\frac{\sin\pi s}\pi\int_{0}^{\infty}h^{-s}\frac{d\log
z(h-i0)}{dh}dh\label{YLaplace}%
\end{equation}
where the contour of integration is shifted from the real axis to agree the
branch chosen in (\ref{zetaYL}). By the way, from (\ref{logzh}) and
(\ref{melline}) it follows that
\begin{equation}
\operatorname*{res}_{s=\rho}\zeta_{\text{YL}}(s)=f_{0}\rho\sin\pi
\rho\label{resro}%
\end{equation}
and
\begin{align}
\zeta_{\text{YL}}(0) &  =\delta+1/2\label{zetinf}\\
\zeta_{\text{YL}}^{\prime}(0) &  =\log z_{\infty}\nonumber
\end{align}

Notice that $\delta$ in (\ref{bzdel}) is always positive (see table
\ref{Table2}). Therefore (\ref{bzdel}) certainly doesn't give any
approximation for the first zero $h_{0}$, which, as it was mentioned above, is
in any case positive. However for the next zeros, even for $h_{1}$ and $h_{2}%
$, asymptotic estimate (\ref{bzdel}) gives surprisingly good approximations.
This curious feature has been found out in the classical case $b^{2}=0$
\cite{sphere} and will be certified now for the quantum case by the numerical
analysis as well as through the exactly solvable case $b^{2}=0.4$ (see below).
This phenomenology motivates the following simple strategy, tried already in
\cite{sphere}: take (\ref{hass}) as a good first approximation for $h_{n}$ at
$n>0$ and try to use the complementary information provided by the
perturbative coefficients (\ref{z123}) to evaluate numerically the unknown
parameter $f_{0}$ and also improve the approximation (\ref{hass}) as far as possible.

To demonstrate this analytic-numeric procedure let's take the simplest version
with only two first coefficients $z_{1}$ and $z_{2}$ taken into account. The
additional restriction for the zeros $h_{n}$ is contained in the sum rules
\begin{align}
\frac1{h_{0}}+\sum_{n=1}^{\infty}\frac1{h_{n}} &  =0\label{rules2}\\
\frac1{h_{0}^{2}}+\sum_{n=1}^{\infty}\frac1{h_{n}^{2}} &  =-2z_{2}\nonumber
\end{align}
Along the above strategy we approximate the sums in (\ref{rules2}) through the
asymptotic (\ref{hass})
\begin{align}
\frac1{h_{0}}-Yr_{1} &  =0\label{eq2}\\
\frac1{h_{0}^{2}}+Y^{2}r_{2} &  =-2z_{2}\nonumber
\end{align}
where $Y=\left(  f_{0}\sin(\pi\rho)\right)  ^{1/\rho}$ and
\begin{align}
r_{1} &  =\zeta(1/\rho,1-\delta)\label{r12}\\
r_{2} &  =\zeta(2/\rho,1-\delta)\nonumber
\end{align}
Here $\zeta(s,a)$ is the usual Riemann zeta function. The system (\ref{eq2})
is easily solved for $h_{0}=(Yr_{1})^{-1}$ and $Y^{2}=-2z_{2}/(r_{2}+r_{1}%
^{2})$ giving the approximation for $f_{0}$%
\begin{equation}
f_{0}^{(2)}=\frac1{\sin(\pi\rho)}\left(  \frac{-2z_{2}}{r_{2}+r_{1}^{2}%
}\right)  ^{\rho/2}\label{f02}%
\end{equation}

If first $N$ perturbative coefficients $z_{1}\ldots,z_{N}$ are known the
corresponding $N$ sum rules (\ref{rules2}) can be solved for $f_{0}$ and
$h_{n}$, $n=0,1,\ldots,N-2$. The corresponding procedure, although quite
simple, is described in appendix A.

Once the parameter $Y$ and the first $N-1$ zeros are restored in this order,
the scaling function $z(h)$ can be evaluated using (\ref{melline}) with the
corresponding approximate $\zeta_{\text{YL}}(s)$
\begin{equation}
\zeta_{\text{YL}}^{(N)}(s)=\zeta_{\text{YL}}(s)=\frac{e^{i\pi s}}{h_{0}^{s}%
}+\sum_{n=1}^{N-2}\frac1{(-h_{n})^{s}}+Y^{s}\zeta(s/\rho,N-1-\delta
)\label{zetaYLN}%
\end{equation}
or
\begin{equation}
z(h)=\prod_{n=0}^{N-2}\left(  1-\frac h{h_{n}}\right)  \exp\left(
\int_{\uparrow}\frac{\pi\zeta(s/\rho,N-1-\delta)}{\sin(\pi s)}(Yh)^{s}%
\frac{ds}{2i\pi s}\right) \label{zhN}%
\end{equation}

\section{Numerics}

In table \ref{Table1} numerical values of the non-trivial perturbative
coefficients $z_{2}$ and $z_{3}$ form (\ref{z123}) are summarized for
different values of $b^{2}$. At certain values of $b^{2}$ the four-point
coefficient is also available and presented in the table.%

\begin{table}[htb] \centering
\begin{tabular}
[c]{|llllll|}\hline
\multicolumn{1}{|l|}{$b^{2}$} & \multicolumn{1}{l|}{$Z_{\text{L}}^{(A)}$} &
\multicolumn{1}{l|}{$z_{2}$} & \multicolumn{1}{l|}{$z_{3}$} &
\multicolumn{1}{l|}{$z_{4}$} & $z_{4}^{\text{(sum)}}$\\\hline
\multicolumn{1}{|l|}{$0.00$} & \multicolumn{1}{l|}{$\infty$} &
\multicolumn{1}{l|}{$-0.0892857$} & \multicolumn{1}{l|}{$0.0229899$} &
\multicolumn{1}{l|}{$-0.0318918$} & $-0.00318903$\\\hline
\multicolumn{1}{|l|}{$0.01$} & \multicolumn{1}{l|}{$1.08\times10^{45}$} &
\multicolumn{1}{l|}{$-0.0883599$} & \multicolumn{1}{l|}{$0.0225977$} &
\multicolumn{1}{l|}{} & $-0.00311130$\\\hline
\multicolumn{1}{|l|}{$0.05$} & \multicolumn{1}{l|}{$7.43\times10^{8}$} &
\multicolumn{1}{l|}{$-0.0843801$} & \multicolumn{1}{l|}{$0.0209364$} &
\multicolumn{1}{l|}{} & $-0.00278803$\\\hline
\multicolumn{1}{|l|}{$0.10$} & \multicolumn{1}{l|}{$7909.85$} &
\multicolumn{1}{l|}{$-0.0786500$} & \multicolumn{1}{l|}{$0.0186240$} &
\multicolumn{1}{l|}{} & $-0.00235567$\\\hline
\multicolumn{1}{|l|}{$0.15$} & \multicolumn{1}{l|}{$116.757$} &
\multicolumn{1}{l|}{$-0.0718331$} & \multicolumn{1}{l|}{$0.0160092$} &
\multicolumn{1}{l|}{} & $-0.00189425$\\\hline
\multicolumn{1}{|l|}{$0.1974$} & \multicolumn{1}{l|}{$12.6331$} &
\multicolumn{1}{l|}{$-0.0640680$} & \multicolumn{1}{l|}{$0.0132252$} &
\multicolumn{1}{l|}{$-0.0014386452$} & $-0.00143865$\\\hline
\multicolumn{1}{|l|}{$0.20$} & \multicolumn{1}{l|}{$11.4592$} &
\multicolumn{1}{l|}{$-0.0635922$} & \multicolumn{1}{l|}{$0.0130616$} &
\multicolumn{1}{l|}{} & $-0.00141311$\\\hline
\multicolumn{1}{|l|}{$0.2404$} & \multicolumn{1}{l|}{$3.19037$} &
\multicolumn{1}{l|}{$-0.0555991$} & \multicolumn{1}{l|}{$0.0104388$} &
\multicolumn{1}{l|}{$-0.0010237712$} & $-0.00102381$\\\hline
\multicolumn{1}{|l|}{$0.25$} & \multicolumn{1}{l|}{$2.47138$} &
\multicolumn{1}{l|}{$-0.0534942$} & \multicolumn{1}{l|}{$0.00978831$} &
\multicolumn{1}{l|}{} & $-0.000933453$\\\hline
\multicolumn{1}{|l|}{$0.30$} & \multicolumn{1}{l|}{$0.799642$} &
\multicolumn{1}{l|}{$-0.0409998$} & \multicolumn{1}{l|}{$0.00628732$} &
\multicolumn{1}{l|}{} & $-0.000495034$\\\hline
\multicolumn{1}{|l|}{$0.35$} & \multicolumn{1}{l|}{$0.327724$} &
\multicolumn{1}{l|}{$-0.0255378$} & \multicolumn{1}{l|}{$0.00287061$} &
\multicolumn{1}{l|}{} & $-0.000161857$\\\hline
\multicolumn{1}{|l|}{$0.40$} & \multicolumn{1}{l|}{$0.155805$} &
\multicolumn{1}{l|}{$-0.00714401$} & \multicolumn{1}{l|}{$0.000356752$} &
\multicolumn{1}{l|}{$-8.50615\times10^{-6}$} & $-8.52261\times10^{-6}$\\\hline
\end{tabular}
\caption{Numerical values for the second, third, and sometimes
forth order perturbative coefficients in the fixed area scaling function $z(h)$.
In the last coloumn we place the estimate of the four-point coefficient from the
sum rules. \label{Table1}}%
\end{table}

\begin{figure}
[tbh]
\begin{center}
\includegraphics[
height=3.6391in,
width=4.67in
]%
{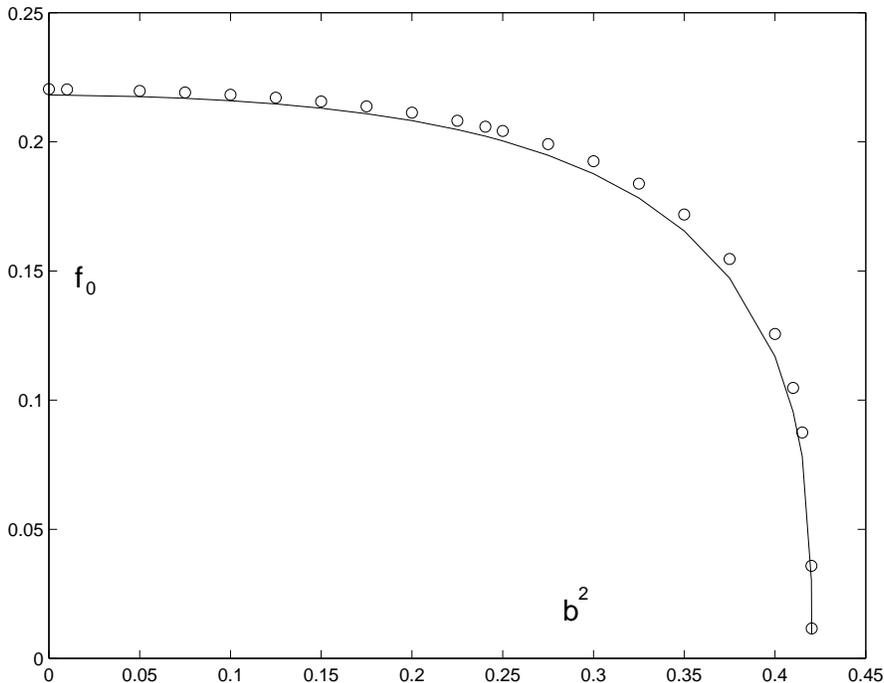}%
\caption{Critical point $f_{0}^{(2)}$ (open circles) and $f_{0}^{(3)}$
(continuous line) as a function of $b^{2}$}%
\label{fig1}%
\end{center}
\end{figure}

Table \ref{Table2} summarizes the results of the analytic-numeric procedure
described in the previous section. Here $f_{0}^{(2)}$, $f_{0}^{(3)}$ and
$f_{0}^{(4)}$ are the estimates of the free energy parameter with the use of
the perturbative coefficients from the table \ref{Table1} up to $z_{2}$,
$z_{3}$ and also $z_{4}$ where the last is available. The exact values
$f_{0}^{\text{(exact)}}$ come from the integrable classical case (\ref{f00})
at $b^{2}=0$ and from the matrix model solution in the case of minimal gravity
$b^{2}=0.4$. These approximate values of the specific gravitational free
energy $f_{0}(b^{2})$ are plotted in fig.\ref{fig1}. Apparently it has a
singularity at the ``critical'' value $b^{2}=b_{\text{c}}^{2}$. This
singularity will be considered in more details elsewhere.%

\begin{table}[htb] \centering
\begin{tabular}
[c]{|lllllll|}\hline
\multicolumn{1}{|l|}{$b^{2}$} & \multicolumn{1}{l|}{$\rho$} &
\multicolumn{1}{l|}{$\delta$} & \multicolumn{1}{l|}{$f_{0}^{(2)}$} &
\multicolumn{1}{l|}{$f_{0}^{(3)}$} & \multicolumn{1}{l|}{$f_{0}^{(4)}$} &
$f_{0}^{\text{(exact)}}$\\\hline
\multicolumn{1}{|l|}{$0.00$} & \multicolumn{1}{l|}{$0.833333$} &
\multicolumn{1}{l|}{$0.111111$} & \multicolumn{1}{l|}{$0.220407$} &
\multicolumn{1}{l|}{$0.218156$} & \multicolumn{1}{l|}{$0.218036$} &
$0.2179745$\\\hline
\multicolumn{1}{|l|}{$0.01$} & \multicolumn{1}{l|}{$0.831646$} &
\multicolumn{1}{l|}{$0.109935$} & \multicolumn{1}{l|}{$0.220318$} &
\multicolumn{1}{l|}{$0.218091$} & \multicolumn{1}{l|}{} & \\\hline
\multicolumn{1}{|l|}{$0.05$} & \multicolumn{1}{l|}{$0.824462$} &
\multicolumn{1}{l|}{$0.106179$} & \multicolumn{1}{l|}{$0.219719$} &
\multicolumn{1}{l|}{$0.217523$} & \multicolumn{1}{l|}{} & \\\hline
\multicolumn{1}{|l|}{$0.10$} & \multicolumn{1}{l|}{$0.814333$} &
\multicolumn{1}{l|}{$0.103698$} & \multicolumn{1}{l|}{$0.218260$} &
\multicolumn{1}{l|}{$0.215950$} & \multicolumn{1}{l|}{} & \\\hline
\multicolumn{1}{|l|}{$0.15$} & \multicolumn{1}{l|}{$0.802587$} &
\multicolumn{1}{l|}{$0.103880$} & \multicolumn{1}{l|}{$0.215654$} &
\multicolumn{1}{l|}{$0.213053$} & \multicolumn{1}{l|}{} & \\\hline
\multicolumn{1}{|l|}{$0.1974$} & \multicolumn{1}{l|}{$0.789474$} &
\multicolumn{1}{l|}{$0.106813$} & \multicolumn{1}{l|}{$0.211594$} &
\multicolumn{1}{l|}{$0.208534$} & \multicolumn{1}{l|}{$0.208240$} & \\\hline
\multicolumn{1}{|l|}{$0.20$} & \multicolumn{1}{l|}{$0.788675$} &
\multicolumn{1}{l|}{$0.107063$} & \multicolumn{1}{l|}{$0.211308$} &
\multicolumn{1}{l|}{$0.208218$} & \multicolumn{1}{l|}{} & \\\hline
\multicolumn{1}{|l|}{$0.2404$} & \multicolumn{1}{l|}{$0.775255$} &
\multicolumn{1}{l|}{$0.112194$} & \multicolumn{1}{l|}{$0.205864$} &
\multicolumn{1}{l|}{$0.202209$} & \multicolumn{1}{l|}{$0.201858$} & \\\hline
\multicolumn{1}{|l|}{$0.25$} & \multicolumn{1}{l|}{$0.771700$} &
\multicolumn{1}{l|}{$0.113797$} & \multicolumn{1}{l|}{$0.204224$} &
\multicolumn{1}{l|}{$0.200408$} & \multicolumn{1}{l|}{} & \\\hline
\multicolumn{1}{|l|}{$0.30$} & \multicolumn{1}{l|}{$0.75$} &
\multicolumn{1}{l|}{$0.125$} & \multicolumn{1}{l|}{$0.192522$} &
\multicolumn{1}{l|}{$0.187677$} & \multicolumn{1}{l|}{} & \\\hline
\multicolumn{1}{|l|}{$0.35$} & \multicolumn{1}{l|}{$0.719788$} &
\multicolumn{1}{l|}{$0.142247$} & \multicolumn{1}{l|}{$0.171896$} &
\multicolumn{1}{l|}{$0.165563$} & \multicolumn{1}{l|}{} & \\\hline
\multicolumn{1}{|l|}{$0.40$} & \multicolumn{1}{l|}{$0.666667$} &
\multicolumn{1}{l|}{$0.166667$} & \multicolumn{1}{l|}{$0.125625$} &
\multicolumn{1}{l|}{$0.116905$} & \multicolumn{1}{l|}{$0.116151$} &
$0.1158596$\\\hline
\end{tabular}
\caption
{Specific energy $f_0$ determined with the use of first two, three and four (when
available) perturbative coefficients.\label{Table2}}%
\end{table}

To illustrate the convergence of the zero locations $h_{n}$ and also how a
good approximation the asymptotic (\ref{hass}) gives even for $h_{1}$, we
collect some related numbers in table \ref{Table3}. The column $h_{1}%
^{\text{(as)}}$ corresponds to $n=1$ in (\ref{hass}) with $f_{0}=f_{0}^{(2)}$
from table \ref{Table2}. The value of $h_{0}^{\text{(exact)}}$ is taken at
$b^{2}=0.4$ from the exact matrix model results (see next section) and at
$b^{2}=0$ as the best TCS fit of \cite{sphere}.%

\begin{table}[htb] \centering
\begin{tabular}
[c]{|l|l|l|l|l|l|l|l|}\hline
$b^{2}$ & $h_{0}^{(2)}$ & $h_{0}^{(3)}$ & $h_{0}^{(4)}$ & $h_{0}%
^{\text{(exact)}}$ & $h_{1}^{\text{(as)}}$ & $h_{1}^{(3)}$ & $h_{1}^{(4)}%
$\\\hline
$0.00$ & $2.42697$ & $2.43068$ & $2.43069$ & $2.43069$ & $-12.2410$ &
$-11.7540$ & $-11.7667$\\\hline
$0.01$ & $2.44074$ & $2.44446$ &  &  & $-12.1991$ & $-11.7193$ & \\\hline
$0.05$ & $2.50261$ & $2.50648$ &  &  & $-12.0120$ & $-11.5656$ & \\\hline
$0.10$ & $2.60000$ & $2.60444$ &  &  & $-11.7880$ & $-11.3549$ & \\\hline
$0.15$ & $2.73096$ & $2.73652$ &  &  & $-11.5990$ & $-11.1519$ & \\\hline
$0.1974$ & $2.90514$ & $2.91258$ & $2.91264$ &  & $-11.4909$ & $-11.0075$ &
$-11.0267$\\\hline
$0.20$ & $2.91685$ & $2.92443$ &  &  & $-11.4880$ & $-11.0020$ & \\\hline
$0.2404$ & $3.13562$ & $3.14598$ & $3.14606$ &  & $-11.5099$ & $-10.9730$ &
$-10.9928$\\\hline
$0.25$ & $3.20130$ & $3.21256$ &  &  & $-11.5386$ & $-10.9867$ & \\\hline
$0.30$ & $3.69093$ & $3.70962$ &  &  & $-11.9506$ & $-11.2907$ & \\\hline
$0.35$ & $4.74483$ & $4.78283$ &  &  & $-13.3915$ & $-12.5134$ & \\\hline
$0.40$ & $9.22903$ & $9.38245$ & $9.38349$ & $9.38350$ & $-21.1991$ &
$-19.3580$ & $-19.3879$\\\hline
\end{tabular}
\caption{Positions of first two zeros of $h(z)$ at different
approximations\label{Table3}}%
\end{table}

To summarize, using only three first coefficients of the power series in
$z(h)$ we have restored to an impressive accuracy the free energy $f_{0}%
^{(3)}$ and also the first two zeros $h_{0}^{(3)}$ and $h_{1}^{(3)}$, the
subsequent being given in this approximation by the asymptotic formula
(\ref{hass}). With this data we can approximately calculate the next
perturbative coefficients through the sum rules similar to (\ref{rules2}).
Such estimate $z_{4}^{\text{(sum)}}$ for the four point coefficient are quoted
in the last column of table \ref{Table1}. These ``extrapolated'' numbers are
compared with the actual ones where the last are available, i.e., at the
classical point \cite{sphere}, the minimal gravity point $b^{2}=0.4$ and, in
addition, at two special values of $b^{2}$ where the integral in the forth
line of (\ref{a1234}) is somewhat simpler and can be carried out (see in sect
11). The comparison is mainly to check the consistency of the four-point
integrals with the expected analytic structure. Another reason is to estimate
the numerical precision to be achieved in the four-point integration of
(\ref{a1234}) in order to add new information to the analytic-numeric data.
The requirements to any numeric implementations turn out to be rather high (in
most cases not less then 5--6 decimal digits of precision is needed).

\section{``Integrable'' point $b^{2}=2/5$}

Point $b^{2}=2/5$ corresponds to $c_{\text{sp}}=0$, i.e., no spectator matter.
The matter content of the critical (unperturbed) gravity is exhausted by a
single minimal CFT model ($\mathcal{M}_{2/5}$ in the present case). This
situation is called the ``minimal gravity''. There are many reasons to believe
that the minimal gravity is solvable, at least in its ``topological'' aspects.
Namely, all (or at least large classes of) integrated multipoint correlation
numbers are in principle calculable. The legend is that some versions of
perturbed minimal gravity, to the utmost radical version \textit{all }possible
perturbations, are related to certain solvable matrix models \cite{matrix}
(more precisely, the classes of their critical behavior). Anyhow, in many
cases this folklore can be promoted to specific statements and our problem of
the perturbed minimal Yang-Lee gravity is among them.

The scaling function $Z_{\text{t}}(x,t)$, which describes the scaling region
near the tricritical point of the one matrix model (at the planar limit
corresponding to the spherical topology), is determined explicitly as
\begin{equation}
\frac{\partial^{2}}{\partial x^{2}}Z_{\text{t}}(x,t)=u(x,t)\label{Zt}%
\end{equation}
through a solution $u(x,t)$ of the following simple algebraic equation
\begin{equation}
x=u^{3}-tu\label{u3}%
\end{equation}
In ref.\cite{Staudacher} this scaling region has been identified with the
perturbed critical $b^{2}=2/5$ minimal gravity, parameters $t$ and $x$ playing
the role of the cosmological constant and $\varphi$-perturbation coupling
respectively. Of course, this identification doesn't fix the overall scale
neither of the partition function (\ref{Zt}) itself nor of the coupling
constants. Thus, we have to identify $t$ and $x$ with $\mu$ and $\lambda$ and
the spherical partition function (\ref{Zscale}) $Z_{\text{t}}(x,t)$ up to
certain normalization constants.

Equations (\ref{Zt}) and (\ref{u3}) result in the following expansion
\begin{align}
Z_{\text{t}}(x,t) &  =t^{7/2}\sum_{n=0}^{\infty}\frac12\frac{\Gamma
(3n/2-7/2)}{n!\Gamma(n/2-1/2)}\left(  \frac x{t^{3/2}}\right)  ^{n}%
\label{Ztexp}\\
&  =Z_{\text{t}}(0,t)\left(  1-\frac{105}{16}\left(  \frac x{t^{3/2}}\right)
+\frac{105}8\left(  \frac x{t^{3/2}}\right)  ^{2}-\frac{35}{16}\left(  \frac
x{t^{3/2}}\right)  ^{3}+\ldots\right) \nonumber
\end{align}
The linear in $x$ term is regular and therefore cannot be described
unambiguously in the field theoretic context. In particular, this term doesn't
contribute to the fixed area partition function below. Hence we do not pay
much attention to this term and instead compare the next terms with the
perturbative coefficients (\ref{a1234}). The $x^{2}$ term leads to the
relation
\begin{equation}
\frac x{t^{3/2}}=\frac{\lambda l_{\text{eg}}}{(\pi\mu)^{3/2}}\label{xtmu}%
\end{equation}
which is also consistent with the $x^{3}$ term. Here
\begin{equation}
l_{\text{eg}}=\frac{\gamma^{1/2}(4/5)}{4\gamma(2/5)}=0.0845223\ldots
\label{leg}%
\end{equation}
Also the singular point in $t$ is determined from (\ref{u3}) and (\ref{xtmu})
\begin{equation}
\mu_{\text{c}}=\frac3\pi\left(  \frac{l_{\text{eg}}}2\right)  ^{2/3}%
=0.11585962159187\label{mc25}%
\end{equation}
This is the number quoted as $f_{0}^{\text{(exact)}}$ in the table
\ref{Table2}.

Neglecting the overall normalization of the partition function, which in any
case doesn't enter the fixed area scaling function $z(h)$, we are in the
position to write the latter explicitly. The coefficients in (\ref{zzn}) are
\begin{equation}
z_{n}=-\frac{2\sqrt{\pi}(-l_{\text{eg}})^{n}}{n!\Gamma(n/2-1/2)}\label{zn25}%
\end{equation}
Numerically
\begin{align}
z_{2} &  =-l_{\text{eg}}^{2}=-0.00714401\nonumber\\
z_{3} &  =\frac{\sqrt{\pi}}3l_{\text{eg}}^{3}=0.000356752\label{znum}\\
z_{4} &  =-\frac16l_{\text{eg}}^{4}=-8.50615\ldots\times10^{-6}\nonumber\\
&  \ldots\nonumber
\end{align}
The last number has been used in the previous section for the numerical analysis.

The following integral representation
\begin{equation}
z(h)=-\frac2{\sqrt{\pi}}\int_{\uparrow}y^{2}dy\exp\left(  y^{2}+\frac
{l_{\text{eg}}h}y\right) \label{zhint}%
\end{equation}
where the integration is parallel to the imaginary axis and goes to the right
from the singularity at $y=0$.

\begin{figure}
[tbh]
\begin{center}
\includegraphics[
height=3.9288in,
width=5.009in
]%
{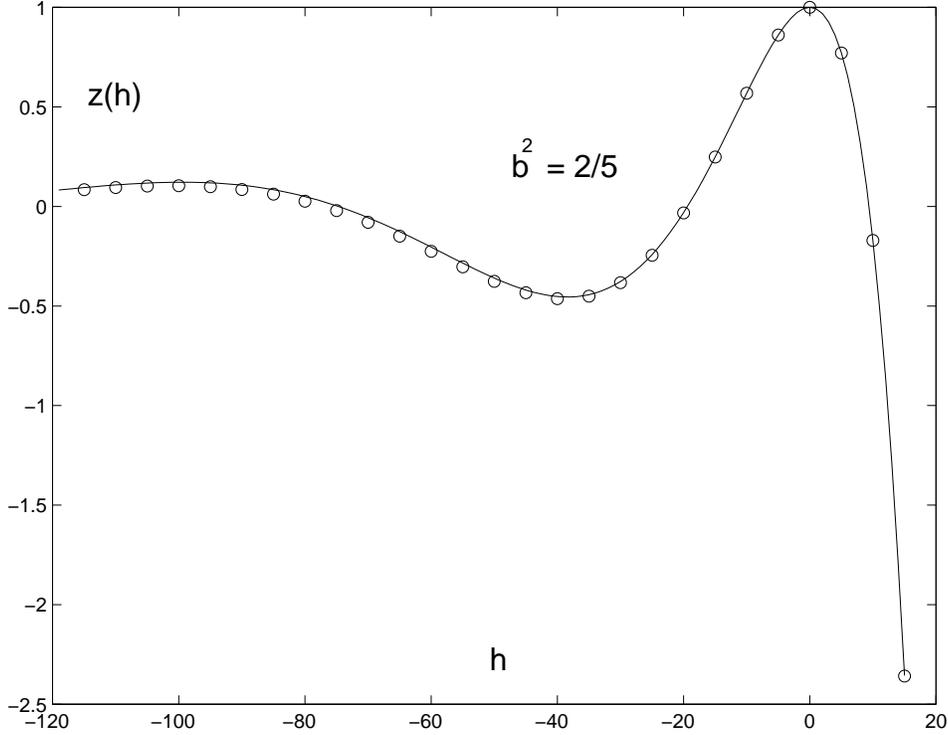}%
\caption{Continuous line: exact function $z(h)$ as defined by (\ref{zhint}).
Open circles: the canonical product over the zeros recovered from $z_{2}$ and
$z_{3}$ through the analytic-numeric procedure}%
\label{fig2}%
\end{center}
\end{figure}

In fig.\ref{fig2} exact scaling function (\ref{zhint}) is plotted. For
comparison we present, without any comments, the numbers obtained by the order
3 (i.e., first three coefficients $z_{1}$, $z_{2}$ and $z_{3}$ are used in the
analysis) analytic-numeric procedure of sect.8.%

It is worth mentioning that the family of gravitational Yang-Lee models $0\leq
b^{2}\leq b_{\text{c}}^{2}$ contains another interesting point $b^{2}=0.3$,
where it can be reduced to the minimal gravity and therefore there are all
reasons to expect the model to be solvable. At this point $c_{\text{sp}%
}=-22/5$ and the model can be thought of as \textit{two independent} Yang-Lee
models, one is perturbed while the other remains conformal (spectator). So far
nothing special, one can make many similar combinations of minimal matter
models constructing different theories of gravity, most of them being
non-solvable (or non-integrable, in a wide sense). Combination $\mathcal{M}%
_{2/5}\oplus\mathcal{M}_{2/5}$ is exceptional, because it admits an
alternative description in terms of \textit{single} minimal model
$\mathcal{M}_{3/10}$ (central charge $-44/5$) \cite{VWI}. The table of
degenerate dimensions of this model%

\begin{equation}%
\begin{tabular}
[c]{|l|l|l|l|l|l|l|l|l|}\hline
$2$ & $49/40$ & $3/5$ & $1/8$ & $-1/5$ & $-3/8$ & $-2/5$ & $-11/40$ &
$0$\\\hline
$0$ & $-11/40$ & $-2/5$ & $-3/8$ & $-1/5$ & $1/8$ & $3/5$ & $49/40$ &
$2$\\\hline
\end{tabular}
\label{M310}%
\end{equation}
includes the fields $\Phi_{1,5}$ and $\Phi_{2,5}$ of dimension $-1/5$. In the
non-diagonal ($D$-series) version of this minimal model certain combination of
these operators is interpreted as $\varphi$ acting only inside one of the
constituent $\mathcal{M}_{2/5}$ models. It is natural to expect that such
perturbed minimal gravity can be related to an appropriate matrix model, and
therefore is solvable. The point seems to merit further study.

\section{Four-point integral}

Relative success of the numerical program of sect.8 makes it tempting to
improve the approximation by involving the next four-order term $z_{4}$. This
coefficient is related to the four-point correlation number. In the framework
of Liouville gravity it requires an integration over moduli (\ref{Un}) of the
product of the matter and the Liouville four-point functions, as in the last
line of eq.(\ref{a1234}). In the fixed area context it is more natural to use
the fixed area Liouville four-point function
\begin{equation}
G_{\text{L}}^{(A=\pi)}(x,\bar x)=\frac{\left.  G_{\text{L}}(x,\bar x)\right|
_{\pi\mu=1}}{\Gamma(4gb^{-1}-Qb^{-1})}\label{G4A}%
\end{equation}
The fixed area term is rewritten as
\begin{equation}
z_{4}=\frac{\mathcal{I}(b)}{24(2\pi)^{4}bZ_{\text{L}}^{(A=\pi)}}\label{z4I}%
\end{equation}
where, as usual
\begin{equation}
Z_{\text{L}}^{(A=\pi)}=\frac{\gamma^{Q/b}(b^{2})\Gamma(2-b^{2})}{\pi^{3}%
b^{3}\Gamma(b^{2})\Gamma(b^{-2})}\label{ZApi}%
\end{equation}
and
\begin{equation}
\mathcal{I}(b)=b\int G_{\text{YL}}(x,\bar x)G_{\text{L}}^{(A=\pi)}(x,\bar
x)d^{2}x\label{Ib}%
\end{equation}
with $G_{\text{YL}}(x,\bar x)$ from eq.(\ref{GYL}).

In general the Liouville four-point function admits the block representation
(\ref{V4block}), which reads in our symmetric case as
\begin{equation}
\left.  G_{\text{L}}(x,\bar x)\right|  _{\pi\mu=1}=\mathcal{R}_{g}\int
^{\prime}\frac{dP}{4\pi}r_{g}(P)\mathcal{F}_{P}\left(  \left.
\begin{array}
[c]{cc}%
\Delta_{g} & \Delta_{g}\\
\Delta_{g} & \Delta_{g}%
\end{array}
\right|  x\right)  \mathcal{F}_{P}\left(  \left.
\begin{array}
[c]{cc}%
\Delta_{g} & \Delta_{g}\\
\Delta_{g} & \Delta_{g}%
\end{array}
\right|  \bar x\right) \label{GL4block}%
\end{equation}
Here the overall factor $\mathcal{R}_{g}$ reads
\begin{equation}
\mathcal{R}_{g}=\left(  \gamma(b^{2})b^{2-2b^{2}}\right)  ^{(Q-4g)/b}%
\frac{\Upsilon_{b}^{4}(b)\Upsilon_{b}^{4}(2g)}{\pi^{2}\Upsilon_{b}%
^{4}(2g-Q/2)}\label{RRg}%
\end{equation}
while the function
\begin{equation}
r_{g}(P)=\frac{\pi^{2}\Upsilon_{b}(2iP)\Upsilon_{b}(-2iP)\Upsilon_{b}%
^{4}(2g-Q/2)}{\Upsilon_{b}^{2}(b)\Upsilon_{b}^{2}(2g-Q/2-iP)\Upsilon_{b}%
^{2}(2g-Q/2+iP)\Upsilon_{b}^{2}(Q/2-iP)\Upsilon_{b}^{2}(Q/2+iP)}\label{rg}%
\end{equation}
admits the following integral representation (convergent at $g>Q/4$)
\begin{equation}
r_{g}(P)=\sinh2\pi b^{-1}P\sinh2\pi bP\exp\left(  -8\int_{0}^{\infty}\frac
{dt}t\frac{\sin^{2}Pt\left(  \cosh^{2}(Q-2g)t-e^{-Qt}\cos^{2}Pt\right)
}{\sinh bt\sinh b^{-1}t}\right) \label{rgint}%
\end{equation}
Integral (\ref{GL4block}) should be taken literally if $Q/2>g>Q/4$, otherwise
there are additional ``discrete'' terms due to the singularities of the
integrand \cite{AAl}. The prime near the integral sign in (\ref{GL4block})
indicates this subtlety.

\begin{figure}
[tbh]
\begin{center}
\includegraphics[
height=4.5299in,
width=4.9234in
]%
{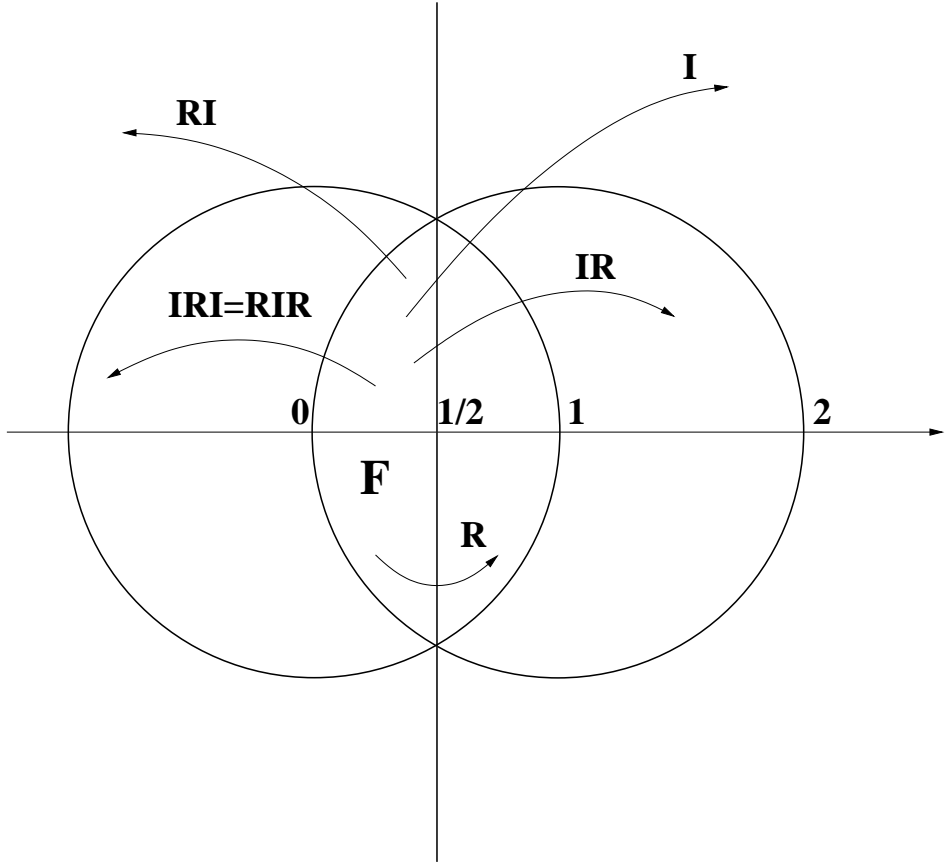}%
\caption{Fundamental region of the modular group (\ref{F}) and the action of
its elements.}%
\label{fig3}%
\end{center}
\end{figure}

Apparently the form $G_{\text{YL}}(x,\bar x)G_{\text{L}}^{(A=\pi)}(x,\bar
x)d^{2}x$ in the integral (\ref{Ib}) is invariant under the group of $6$
modular transformations
\begin{align}
R &  :x\rightarrow1-x\nonumber\\
I &  :x\rightarrow1/x\nonumber\\
RI &  :x\rightarrow1/(1-x)\label{F}\\
IR &  :x\rightarrow1-1/x\nonumber\\
IRI=RIR &  :x\rightarrow x/(x-1)\nonumber
\end{align}
Therefore the integration in (\ref{Ib}) can be restricted to its fundamental
domain $\mathbf{F=}\{{\operatorname*{Re}}x<1/2;\;\left|  1-x\right|  <1\}$
(see fig.\ref{fig3})
\begin{equation}
\mathcal{I}(b)=6b\int_{\mathbf{F}}G_{\text{YL}}(x,\bar x)G_{\text{L}}%
^{(A=\pi)}(x,\bar x)d^{2}x\label{IbF}%
\end{equation}%

Evaluation of integrals in (\ref{Ib}) or (\ref{IbF}), even numerical, presents
certain technical difficulties. They are mostly related to the complicated
nature of the Liouville four point function. Expression (\ref{GL4block})
involves additional integration in (\ref{rgint}) to evaluate the special
functions. This makes the integral in (\ref{IbF}) three-fold. In addition the
general conformal block entering (\ref{GL4block}) is usually known only as a
power series in $x$, its evaluation requiring certain recursive procedure
\cite{Block}.

There is another representation of integral (\ref{Ib}) which takes advantage
of the holomorphic factorization in each term of the matter and Liouville
block decompositions (\ref{GYL}) and (\ref{V4block}) respectively. It is based
on the following simple calculation. Consider the integral
\begin{equation}
\mathcal{I}=\int G(x,\bar x)d^{2}x\label{Iexamp}%
\end{equation}
where $G(x,\bar x)$ is singular at $x=0$, $1$ and $\infty$ only and admits the
representation
\begin{equation}
G(x,\bar x)=\sum_{i}A_{i}\mathcal{F}_{i}(x)\mathcal{F}_{i}(\bar x)\label{Ghah}%
\end{equation}
(the sum over $i$ might be as well an integral over a continuous spectrum or
both) with some coefficients $A_{i}$. The holomorphic ``blocks'' are singular
at $x=0$, where $\mathcal{F}_{i}(x)\sim x^{\Delta_{i}}$, and also at $x=1$ and
$x=\infty$, the combination (\ref{Ghah}) being arranged so as to form a
single-valued function of $x$. Straightforward application of the Stokes
theorem gives
\begin{equation}
\mathcal{I}=\frac1{2i}\sum_{i}A_{i}I_{i}J_{i}\label{GIJ}%
\end{equation}
where
\begin{align}
I_{i}  & =\int_{0}^{1}\mathcal{F}_{i}(x)dx\label{IJ}\\
J_{i}  & =\int_{C}\mathcal{F}_{i}(x)dx\nonumber
\end{align}
and the contour $C$ goes from $-\infty$ to $-\infty$ around $x=0$
counterclockwise. If, in addition the integrand in (\ref{Iexamp}) is symmetric
with respect to the modular group (\ref{F}), equation (\ref{GIJ}) is reduced
to
\begin{equation}
\mathcal{I}=-\sum_{i}\sin(\pi\Delta_{i})A_{i}I_{i}^{2}\label{GII}%
\end{equation}

In our particular problem this identity reads (again up to certain discrete
terms at $g<Q/4$)
\begin{equation}
I(b)=-b\mathcal{R}_{g}\int^{\prime}\frac{dP}{4\pi}r_{g}(P)\left(  \sin
\pi\Delta_{P}I_{I}^{2}(P)-\kappa^{2}\sin\pi(\Delta_{P}-1/5)I_{\varphi}%
^{2}(P)\right) \label{IbYL}%
\end{equation}
where
\begin{align}
I_{I}(P) &  =\int_{0}^{1}\mathcal{F}_{I}(x)\mathcal{F}_{P}\left(  \left.
\begin{array}
[c]{cc}%
\Delta_{g} & \Delta_{g}\\
\Delta_{g} & \Delta_{g}%
\end{array}
\right|  x\right)  dx\label{IIIphi}\\
I_{\varphi}(P) &  =\int_{0}^{1}\mathcal{F}_{\varphi}(x)\mathcal{F}_{P}\left(
\left.
\begin{array}
[c]{cc}%
\Delta_{g} & \Delta_{g}\\
\Delta_{g} & \Delta_{g}%
\end{array}
\right|  x\right)  dx\nonumber
\end{align}
$\mathcal{F}_{I}(x)$ and $\mathcal{F}_{\varphi}(x)$ are from
eq.(\ref{YLblocks}).

Expression (\ref{IbYL}) has an obvious advantage compared with the form
(\ref{IbF}), since it involves one integration less. However, evaluation of
the line integrals (\ref{IIIphi}) requires the Liouville block to be computed
to the required accuracy up to the upper limit $x=1$. For a function known
mainly through its power series this makes an additional problem. On the other
hand, representation (\ref{IbF}) contains only the blocks with $\left|
x\right|  <\sqrt{3}/2$, where the convergence of the block power series is
still reasonable. To conclude, it is difficult to say in advance which one of
these two representations is better for numerical work. In the next section we
take two special values of $b^{2}$ where the Liouville correlation function
can be found in closed form. The Liouville blocks explicitly known make
formula (\ref{IbYL}) very effective and accurate. This will allow us to
estimate the numerical precision of the direct integration in (\ref{IbF}).

\section{Special $b^{2}$}

Liouville correlation function $\left\langle e^{2a_{1}\phi(x_{1})}\ldots
e^{2a_{N}\phi(x_{N})}\right\rangle _{\text{L}}$ develop poles if its
parameters $a_{i}$ satisfy certain relations. Following A.Polyakov \cite{Res}
we will call them the ``on-mass-shell'' conditions and the corresponding
correlations the ``resonant'' ones. The simplest on-mass-shell condition reads
for $a=\sum_{i=1}^{N}a_{i}$
\begin{equation}
a=Q-nb\label{anres}%
\end{equation}
with $n$ non-negative integer. It can be easily traced to the divergency in
the integration over the ``zero mode'' of the Liouville field \cite{GoulianLi,
DiFrancescoKut}. The residues in the resonant poles (\ref{anres}) can be
``naively'' red off from the Liouville Lagrangian (\ref{LLFT})
\begin{equation}
\left\langle e^{2a_{1}\phi(x_{1})}\ldots e^{2a_{N}\phi(x_{N})}\right\rangle
_{\text{L}}\sim\frac{(-\mu)^{n}}{n!}\frac{\mathcal{G}_{n}(a_{1},\ldots,a_{N}%
)}{a-Q+nb}\label{nresonance}%
\end{equation}
where
\begin{equation}
\mathcal{G}_{n}(a_{1},\ldots,a_{n})=\int\left\langle e^{2a_{1}\phi(x_{1}%
)}\ldots e^{2a_{n}\phi(x_{n})}\prod_{k=1}^{n}e^{2b\phi(y_{k})}d^{2}%
y_{k}\right\rangle \label{Gnres}%
\end{equation}
and the expectation value in the right hand side is over free massless boson
without the zero mode.

In particular in the four-point case the simplest $n=0$ resonance doesn't
require any integrals in the right hand side and the ``reduced'' four point
function of (\ref{V4}) takes the form
\begin{equation}
G_{\text{L}}\left(  \left.
\begin{array}
[c]{cc}%
a_{1} & a_{3}\\
a_{2} & a_{4}%
\end{array}
\right|  x,\bar x\right)  =\frac{\mathcal{G}_{0}(x,\bar x)}{a-Q}\label{G0res}%
\end{equation}
where
\begin{equation}
\mathcal{G}_{0}(x,\bar x)=(x\bar x)^{-2a_{1}a_{2}}((1-x)(1-\bar x))^{-2a_{1}%
a_{3}}\label{GG0}%
\end{equation}
Similarly, the next resonance at $a=Q-b$ is controlled by a single ``screening
integral'' in (\ref{Gnres})
\begin{equation}
\mathcal{G}_{1}(x,\bar x)=\frac1{(x\bar x)^{2a_{1}a_{2}}((1-x)(1-\bar
x))^{2a_{1}a_{3}}}\int\frac{d^{2}y}{(y\bar y)^{2a_{2}b}((1-y)(1-\bar
y))^{2a_{3}b}((x-y)(\bar x-\bar y))^{2a_{1}b}}\label{integral}%
\end{equation}
It admits the two-term holomorphic decomposition
\begin{equation}
\mathcal{G}_{1}(x,\bar x)=k_{1}\mathcal{F}_{1}(x)\mathcal{F}_{1}(\bar
x)+k_{2}\mathcal{F}_{2}(x)\mathcal{F}_{2}(\bar x)\label{GG1}%
\end{equation}
where
\begin{align}
k_{1} &  =\frac{\pi\gamma(1-2a_{2}b-2a_{1}b)\gamma(1-2a_{3}b)}{\gamma
(2-2a_{2}b-2a_{3}b-2a_{1}b)}\label{a12}\\
k_{2} &  =\frac{\pi\gamma(1-2a_{2}b)\gamma(1-2a_{1}b)}{\gamma(2-2a_{1}%
b-2a_{2}b)}\nonumber
\end{align}
and
\begin{align}
\mathcal{F}_{1}(x)  & =x^{-2a_{1}a_{2}}(1-x)^{-2a_{1}a_{3}}{}{}_{2}%
F_{1}(2a_{1}b+2a_{2}b+2a_{3}b-1,2a_{1}b,2a_{1}b+2a_{2}b,x)\label{F12hyp}\\
\mathcal{F}_{2}(x)  & =x^{1-2a_{2}b-2a_{1}b-2a_{1}a_{2}}(1-x)^{-2a_{1}a_{3}}%
{}_{2}F_{1}(1-2a_{2}b,2a_{3}b,2-2a_{1}b-2a_{2}b,x)\nonumber
\end{align}
To check the consistency of the resonance condition in LFT, in Appendix B we
rederive the residues (\ref{GG0}) and (\ref{GG1}) from the general four-point
expression (\ref{V4block}).

From eq.(\ref{G4A}) it is clear that the poles at (\ref{anres}) are canceled
out in the fixed area correlation functions and the finite parts
\begin{equation}
\left\langle e^{2a_{1}\phi(x_{1})}\ldots e^{2a_{N}\phi(x_{N})}\right\rangle
_{\text{L}}^{(A)}=b^{-1}A^{-n}\mathcal{G}_{n}(a_{1},\ldots,a_{N})\label{AGGn}%
\end{equation}
are expressed through (\ref{Gnres}). In our four-point case this means that
(\ref{Ib}) reads
\begin{equation}
\mathcal{I}(b_{n})=\frac1{\pi^{n}}\int G_{\text{YL}}(x,\bar x)\mathcal{G}%
_{n}(x,\bar x)d^{2}x\label{Ibn}%
\end{equation}

\textbf{1. The first resonant point }$n=0$ appears in our particular problem
(\ref{G0res}) at $b=b_{0}$
\begin{equation}
b_{0}^{2}=\frac{11-4\sqrt{6}}5=0.240408205773\ldots\label{b0}%
\end{equation}
At this point the Liouville dressing parameter $g^{2}=2/5$ so that
\begin{equation}
\mathcal{G}_{0}(x,\bar x)=\left[  x\bar x(1-x)(1-\bar x)\right]
^{-4/5}\label{GG025}%
\end{equation}
Hence
\begin{equation}
\mathcal{I}(b_{0})=\int\frac{\mathcal{F}_{I}(x)\mathcal{F}_{I}(\bar
x)-\kappa^{2}\mathcal{F}_{\varphi}(x)\mathcal{F}_{\varphi}(\bar x)}{\left[
x\bar x(1-x)(1-\bar x)\right]  ^{4/5}}d^{2}x\label{Ib0int}%
\end{equation}
Direct integration over the fundamental region gives (\texttt{NIntegrate} of
Mathematica 3.0)
\begin{equation}
\mathcal{I}(b_{0})=6\int_{\mathbf{F}}\frac{G_{\text{YL}}(x,\bar x)}{\left[
x\bar x(1-x)(1-\bar x)\right]  ^{4/5}}d^{2}x=-59.9029678...\label{Ib0dir}%
\end{equation}

The reduction formula (\ref{GII}) gives in this case two terms
\begin{equation}
\mathcal{I}(b_{0})=\sin\frac{2\pi}5I_{1}^{2}-\kappa^{2}\sin\frac{3\pi}%
5I_{2}^{2}\label{I0sum}%
\end{equation}
Here both contour integrals are carried out explicitly
\begin{align}
I_{1} &  =%
{\displaystyle\int_{0}^{1}}
\frac{F(2/5,3/5,6/5,t)}{t^{2/5}(1-t)^{3/5}}dt=\frac{2^{4/5}\pi(\sqrt
{5}-1)\Gamma^{2}(6/5)}{\Gamma(7/10)\Gamma(9/10)\Gamma(9/5)}\label{I12gamma}\\
I_{2} &  =%
{\displaystyle\int_{0}^{1}}
\frac{F(1/5,2/5,4/5,t)}{t^{3/5}(1-t)^{3/5}}dt=\frac{\sqrt{\pi}\Gamma
(1/5)}{2^{2/5}\Gamma(7/10)}\nonumber
\end{align}
and (\ref{I0sum}) sums up to
\begin{equation}
\mathcal{I}(b_{0})=-\frac\pi{50}\gamma^{5}(1/5)=-59.9030606...\label{I0gamma}%
\end{equation}
Finally
\begin{equation}
z_{4}=\frac{\mathcal{I}(b_{0})}{384b\pi^{4}Z_{\text{L}}^{(A)}}%
=-0.00102377123\ldots\label{z40}%
\end{equation}
is the number used in sect.8 for this special point.

\textbf{2. The second resonant point }$n=1$ corresponds to
\begin{align}
b_{1}^{2}  & =\frac{15}{76}=0.197368421\ldots\label{b1}\\
g^{2}  & =\frac{19}{60}\nonumber
\end{align}
Notice that $gb=1/4$ and thus the integral (\ref{integral}) reads
\begin{equation}
\mathcal{G}_{1}(x,\bar x)=\frac1{[(x\bar x)(1-x)(1-\bar x)]^{2g^{2}}}\int
\frac{d^{2}y}{\left[  y\bar y(1-y)(1-\bar y)(x-y)(\bar x-\bar y)\right]
^{1/2}}\label{G1b1}%
\end{equation}
This fact is general for the $n=1$ resonance in the symmetric case, for
$\sum_{i=1}^{4}a_{i}=4g=Q-b$. The integral is a degenerate case of
(\ref{integral}). The result is recovered at the limit $2a_{i}b\rightarrow1/2$
in (\ref{GG1})
\begin{equation}
\mathcal{G}_{1}(x,\bar x)=\frac{4\left[  K(x)K(1-\bar x)+K(1-x)K(\bar
x)\right]  }{[x\bar x(1-x)(1-\bar x)]^{19/30}}\label{G1K}%
\end{equation}
in terms of the elliptic integral of the first kind
\begin{equation}
K(x)=\frac12\int_{0}^{1}\frac{dy}{\sqrt{y(1-y)(1-xy)}}\label{K}%
\end{equation}
Collecting all together we find
\begin{equation}
\mathcal{I}(b_{1})=\frac1\pi\int G_{\text{YL}}(x,\bar x)\mathcal{G}_{1}(x,\bar
x)d^{2}x\label{Ib1}%
\end{equation}

Direct integration gives with Mathematica 3.0
\begin{equation}
\mathcal{I}(b_{1})=\frac6\pi\int_{\mathbf{F}}G_{\text{YL}}(x,\bar
x)\mathcal{G}_{1}(x,\bar x)d^{2}x=-302.01932532722\ldots\label{Ib1num}%
\end{equation}
The reduction formula (\ref{GII}) doesn't apply here directly due to the
degeneration. It is more convenient to reduce first a more general integral
with (\ref{integral}) as the Liouville part. Then the limit $2a_{i}%
b\rightarrow1/2$ reveals
\begin{equation}
\mathcal{I}(b_{1})=\frac8\pi\left(  \sin\frac{7\pi}{30}I_{I}J_{I}-2\kappa
^{2}\sin\frac{13\pi}{30}I_{\varphi}J_{\varphi}\right) \label{Ib1IJ}%
\end{equation}
where
\begin{equation}
I_{I,\varphi}=\int_{0}^{1}\frac{\mathcal{F}_{I,\varphi}(x)K(x)}%
{(x(1-x))^{19/30}}dx\;;\;\;\;J_{I,\varphi}=\int_{0}^{1}\frac{\mathcal{F}%
_{I,\varphi}(x)K(1-x)}{(x(1-x))^{19/30}}dx\label{IJint}%
\end{equation}
In numbers
\begin{equation}
\mathcal{I}(b_{1})=-302.01932904124\ldots\label{Ib1exact}%
\end{equation}
and%

\begin{equation}
z_{4}=\frac{\mathcal{I}(b_{1})}{384b\pi^{4}Z_{\text{L}}^{(A)}}%
=-0.0014386452412875\ldots\label{z4b1}%
\end{equation}

Reduction formula gives more accurate numbers. This permits to estimate the
precision of the direct integration. In the above two examples it varies
between 5--8 decimal digits. The accuracy of the direct integration is
determined mainly by the singular behavior near $x=0$ in the fundamental
region. Thus, for $b^{2}$ closer to the critical point $b_{\text{c}}^{2}$,
where the ``tachyon'' divergence appears, the direct integration is hardly
expected to give a good approximation.

The following remark is in order here. At the fixed area picture the $n$-th
resonance (\ref{anres}) correlation function gives in the partition function a
contribution proportional to the negative integer powers $A^{-n}$. The
transform to the grand partition function (\ref{Laplace}) results in the terms
containing $\mu^{n}\log\mu$. This is quite usual phenomenon in the field
theory: integer powers of the coupling constant are usually decorated by
logarithm. Otherwise these terms are regular in $\mu$, i.e. contact (from the
common point of view they are out of the field theory scope). On the other
hand, the analysis of the critical behavior in solvable matrix models is
similar to that in the mean field, or Ginzburg-Landau theory, and thus deals
only with the algebraic functions where there is no place for the logarithms.
This might give an impression that the mean field like pattern of critical
behavior is common for the statistical systems interacting with quantum
gravity. We have seen in the above examples that this is certainly not the
case and might be true at most for the systems related to the solvable matrix models.

\section{Discussion}

I think that the main lesson to be learned from those rather random
calculations presented above is that the perturbed Liouville gravity based on
the existing LFT constructions is indeed a consistent tool to treat the
standard problems in 2D gravity. Although at present it cannot yet give any
exact description even in the problems known to be exactly solvable, sometimes
it works equally well in the situations which are not covered by the solvable
matrix models and therefore most probably are not exactly solvable. On the
other hand, non-solvable gravities certainly exist and might be observed e.g.,
as critical points of non-solvable random lattice statistical systems. More
generally, it is likely too hasty to claim the field theory obsolete,
substituted by the matrix models (or whatsoever).

As for the ``spectator matter'' appeared in an apparently artificial way,
there are many practical ways to introduce such content in the statistical
models of lattice gravity. The simplest way is to add a $D$-dimensional free
massless lattice boson (without zero mode) which doesn't interact with other
matter degrees of freedom and contributes simply as an extra weight factor
$\det^{-D/2}\Delta_{\text{lat}}$, where $\Delta_{\text{lat}}$ is the discrete
Laplace operator of the graph. Apparently this parameter $D=c_{\text{sp}}$ can
be as well made continuous. Not to talk about the obvious idea of putting
several extra non-interacting spin-like degrees of freedom and tuning them to
their critical points. As far as I understand, all these situations can be
studied either through the extrapolation of the finite lattice ensembles or in
the Monte-Carlo simulations.

\textbf{Acknowledgments}

My special gratitude is to Galina Gritsenko for her permanent moral support
during all these years of gradual composing the manuscript. The writing has
been finished while the author stayed with a visit at Kawai Theoretical
Laboratory of RIKEN. The hospitality, stimulating scientific atmosphere of the
group and discussions with Y.Ishimoto are highly acknowledged. I am also
obliged to Alexander Zamolodchikov for his authentic and intelligent interest
to the work. Different parts of this study has been reported many times at
different seminars. I thank all my friends and colleagues for their sincere
attempts to understand the motivations to mess around with this ugly
non-integrable pet of mine. The work was supported by the European Committee
under contract EUCLID HRPN-CT-2002-00325.

\appendix

\section{Series analysis}

Suppose that the order $\rho$ of $z(h)$ is less then $1$ and therefore
\begin{equation}
z(h)=\prod_{n=0}^{\infty}\left(  1-\frac h{h_{n}}\right)  =\prod_{n=0}%
^{N-2}\left(  1-\frac h{h_{n}}\right)  \exp\left(  -\sum_{k=1}^{\infty}%
r_{k}(-h)^{k}\right) \label{zexp}%
\end{equation}
where $N$ is any natural number (for our particular problem $N\geq2$) and
\begin{equation}
r_{k}=\frac{(-1)^{k}}k\sum_{N-1}^{\infty}\frac1{h_{n}^{k}}\label{rk}%
\end{equation}
Thus
\begin{equation}
P_{N}(h)=z(h)\exp\left(  \sum_{k=1}^{\infty}r_{k}(-h)^{k}\right) \label{PN}%
\end{equation}
is a polynomial of order $N-1$ with zeros at $h_{n}$, $n=0,1,\ldots,N-2$ (we
have introduced certain unnecessary minus signs in the notations for further convenience).

This set of trivial identities gives rise to the following algorithm for the
problem of sect.7 : solve first truncated $N$ sum rules (\ref{rules2}) for $Y
$ and first $N-1$ zeros $h_{0},\ldots,h_{N-2}$ given the first $N$ non-trivial
coefficients $z_{n}$ in
\begin{equation}
z(h)=\sum_{n=0}^{\infty}z_{n}(-h)^{n}\label{zhagain}%
\end{equation}
The truncation means that all terms in the r.h.s. of (\ref{rules2}) with
$n>N-2$ are replaced by the corresponding asymptotic values (\ref{hass}). In
our present algebraic context the coefficients $r_{k}$ in (\ref{rk}) are taken
as
\begin{equation}
r_{k}=d_{k}Y^{k}\label{rkY}%
\end{equation}
with some known numbers $d_{k}$ (in our actual context $d_{k}=k^{-1}%
\zeta(k/\rho,N-1-\delta)$). Then
\begin{equation}
\exp\left(  \sum_{k=1}^{\infty}r_{k}(-h)^{k}\right)  =\sum_{n=0}^{\infty}%
a_{n}(-hY)^{n}\label{expa}%
\end{equation}
is a regular series in $hY$ with known coefficients
\begin{equation}
a_{n}=\sum_{k_{1}+\ldots+k_{l}=n}\prod_{l}\frac{d_{l}^{k_{l}}}{k_{l}%
!}\label{anrk}%
\end{equation}
The coefficient of order $N$ in (\ref{PN}) is zero. This gives the algebraic
equation for $Y$%
\begin{equation}
Q_{N}(Y)=\sum_{k=0}^{N}a_{k}z_{N-k}Y^{k}=0\label{QN}%
\end{equation}
Once $Y$ is found as a suitable solution to this equation, the $N-1$ roots of
the polynomial are found as the roots of the polynomial
\begin{equation}
\sum_{n=0}^{N-1}Q_{n}(Y)(-h)^{n}=0\label{halg}%
\end{equation}
where
\begin{equation}
Q_{n}(Y)=\sum_{k=0}^{n}a_{k}z_{n-k}Y^{k}\label{Qn}%
\end{equation}

\section{Resonant Liouville correlations}

Consider the general Liouville four-point function (\ref{V4block})
\[
\ \int_{\uparrow}\frac{dc}{4\pi i}C_{\text{L}}\left(  a_{1},a_{2},c\right)
C_{\text{L}}\left(  Q-c,a_{3},a_{4}\right)  \mathcal{F}\left(  \left.
\begin{array}
[c]{cc}%
\Delta_{1} & \Delta_{3}\\
\Delta_{2} & \Delta_{4}%
\end{array}
\right|  \Delta_{c};x\right)  \mathcal{F}\left(  \left.
\begin{array}
[c]{cc}%
\Delta_{1} & \Delta_{3}\\
\Delta_{2} & \Delta_{4}%
\end{array}
\right|  \Delta_{c};\bar x\right)
\]
in the vicinity of the first resonant point
\begin{equation}
a_{1}+a_{2}+a_{3}+a_{4}-Q=\epsilon\label{res0eps}%
\end{equation}
with $\epsilon$ small. Due to the denominators $\Upsilon_{b}(a_{1}+a_{2}-c)$
and $\Upsilon_{b}(a_{3}+a_{4}+c-Q)$ in
\begin{align}
\  & C_{\text{L}}(a_{1},a_{2},c)C_{\text{L}}(Q-c,a_{3},a_{4})=\left(  \pi
\mu\gamma(b^{2})b^{2-2b^{2}}\right)  ^{(Q-a_{1}-a_{2}-a_{3}-a_{4})/b}%
\times\nonumber\\
& \ \ \ \ \ \frac{\Upsilon_{b}(b)\Upsilon_{b}(2a_{1})\Upsilon_{b}%
(2a_{2})\Upsilon_{b}(2c)}{\Upsilon_{b}(a_{1}+a_{2}+c-Q)\Upsilon_{b}%
(a_{1}+a_{2}-c)\Upsilon_{b}(c+a_{1}-a_{2})\Upsilon_{b}(c+a_{2}-a_{1})}%
\times\label{CLCL}\\
& \ \ \ \ \ \frac{\Upsilon_{b}(b)\Upsilon_{b}(2a_{3})\Upsilon_{b}%
(2a_{4})\Upsilon_{b}(2Q-2c)}{\Upsilon_{b}(a_{3}+a_{4}+c-Q)\Upsilon_{b}%
(a_{3}+a_{4}-c)\Upsilon_{b}(c+a_{3}-a_{4})\Upsilon_{b}(c+a_{4}-a_{3}%
)}\nonumber
\end{align}
there are close poles: at $c=$ $a_{1}+a_{2}-\epsilon$ to the left from the
integration contour and at $c=a_{1}+a_{2}$ to the right. This pinch results in
the singular contribution
\begin{equation}
\frac1{2\epsilon}\mathcal{F}\left(  \left.
\begin{array}
[c]{cc}%
\Delta_{1} & \Delta_{3}\\
\Delta_{2} & \Delta_{4}%
\end{array}
\right|  \Delta_{a_{1}+a_{2}};x\right)  \mathcal{F}\left(  \left.
\begin{array}
[c]{cc}%
\Delta_{1} & \Delta_{3}\\
\Delta_{2} & \Delta_{4}%
\end{array}
\right|  \Delta_{a_{1}+a_{2}};\bar x\right) \label{half0}%
\end{equation}
All other multipliers in (\ref{CLCL}) cancel out in the residue. It is not a
problem to verify that
\begin{equation}
\mathcal{F}\left(  \left.
\begin{array}
[c]{cc}%
\Delta_{1} & \Delta_{3}\\
\Delta_{2} & \Delta_{4}%
\end{array}
\right|  \Delta_{a_{1}+a_{2}};x\right)  =x^{-2a_{1}a_{2}}(1-x)^{-2a_{1}a_{3}%
}\label{block0}%
\end{equation}
A ``mirror'' contribution comes from the pinch at $c=Q-a_{1}-a_{2}$ provided
by the $\Upsilon_{b}(a_{1}+a_{2}+c-Q)$ and $\Upsilon_{b}(a_{3}+a_{4}-c)$
denominators in (\ref{CLCL}). Its contribution is identical and we arrive at
(\ref{GG0}). Of course the same situation appears if one or more of $a_{i}$ in
(\ref{res0eps}) are reflected, e.g., $a_{1}+a_{2}+a_{3}-a_{4}=\epsilon$ or
$Q+a_{1}-a_{2}+a_{3}-a_{4}=\epsilon$. Then other pairs of $\Upsilon_{b}%
$-functions give rise to the singularity, which remains the same up to the
expected reflection factors.

Let's now take the second resonant point
\begin{equation}
a_{1}+a_{2}+a_{3}+a_{4}-Q=-b+\epsilon\label{res1eps}%
\end{equation}
The same pair of denominators $\Upsilon_{b}(a_{1}+a_{2}-c)$ and $\Upsilon
_{b}(a_{3}+a_{4}+c-Q)$ in (\ref{CLCL}) produces two pairs of close poles at
$c_{1}=$ $a_{1}+a_{2},\;a_{1}+a_{2}-\epsilon$ and $c_{2}=a_{1}+a_{2}%
+b,\;a_{1}+a_{2}+b-\epsilon$. The singular residues are
\begin{equation}
-\frac{\pi\mu}{2\epsilon}\frac{\gamma(1-2a_{1}b-2a_{2}b)}{\gamma
(2a_{3}b)\gamma(2a_{4}b)}\mathcal{F}\left(  \left.
\begin{array}
[c]{cc}%
\Delta_{1} & \Delta_{3}\\
\Delta_{2} & \Delta_{4}%
\end{array}
\right|  \Delta_{a_{1}+a_{2}};x\right)  \mathcal{F}\left(  \left.
\begin{array}
[c]{cc}%
\Delta_{1} & \Delta_{3}\\
\Delta_{2} & \Delta_{4}%
\end{array}
\right|  \Delta_{a_{1}+a_{2}};\bar x\right) \label{term1}%
\end{equation}
and
\begin{equation}
-\frac{\pi\mu}{2\epsilon}\frac{\gamma(1-2a_{3}b-2a_{3}b)}{\gamma
(2a_{2}b)\gamma(2a_{1}b)}\mathcal{F}\left(  \left.
\begin{array}
[c]{cc}%
\Delta_{1} & \Delta_{3}\\
\Delta_{2} & \Delta_{4}%
\end{array}
\right|  \Delta_{a_{1}+a_{2}+b};x\right)  \mathcal{F}\left(  \left.
\begin{array}
[c]{cc}%
\Delta_{1} & \Delta_{3}\\
\Delta_{2} & \Delta_{4}%
\end{array}
\right|  \Delta_{a_{1}+a_{2}+b};\bar x\right) \label{term2}%
\end{equation}
The ``mirror'' pair $\Upsilon_{b}(a_{1}+a_{2}+c-Q)$ and $\Upsilon_{b}%
(a_{3}+a_{4}-c)$ contributes identically (in fact, this is a general feature
related to the symmetries of the integrand in (\ref{V4block})). The identities
for the blocks
\begin{align}
\  & \mathcal{F}\left(  \left.
\begin{array}
[c]{cc}%
\Delta_{1} & \Delta_{3}\\
\Delta_{2} & \Delta_{4}%
\end{array}
\right|  \Delta_{a_{1}+a_{2}};x\right)  =\label{block1}\\
& \ \ \ x^{-2a_{1}a_{2}}(1-x)^{-2a_{1}a_{3}}{}{}_{2}F_{1}(2a_{1}%
b+2a_{2}b+2a_{3}b-1,2a_{1}b,2a_{1}b+2a_{2}b,x)\nonumber
\end{align}
and
\begin{align}
\  & \ \mathcal{F}\left(  \left.
\begin{array}
[c]{cc}%
\Delta_{1} & \Delta_{3}\\
\Delta_{2} & \Delta_{4}%
\end{array}
\right|  \Delta_{a_{1}+a_{2}+b};x\right)  =\label{block2}\\
& \ \ \ x^{1-2a_{2}b-2a_{1}b-2a_{1}a_{2}}(1-x)^{-2a_{1}a_{3}}{}_{2}%
F_{1}(1-2a_{2}b,2a_{3}b,2-2a_{1}b-2a_{2}b,x)\nonumber
\end{align}
bring us to eq.(\ref{GG1}).

\end{document}